\documentclass[a4paper,11pt]{article}
\pdfoutput=1  
\usepackage[utf8]{inputenc}
\usepackage{jheppub}

\usepackage{amssymb}
\usepackage{physics}
\usepackage{graphicx}
\usepackage{hyperref}
\usepackage{slashed}
\usepackage{bbm}

\newcommand{\be}{\begin{equation}}
\newcommand{\ee}{\end{equation}}
\newcommand{\bea}{\begin{eqnarray}}
\newcommand{\eea}{\end{eqnarray}}

\def \del{\partial}
\def\bea{\begin{eqnarray}}
\def\eea{\end{eqnarray}}
\newcommand{\nn}{\nonumber}
\def\htt{h^{\textmd{TT}}}
\def \pt{\partial}

\begin{document}
 \title{
The weak-gravity bound and the need for spin in asymptotically safe matter-gravity models
 }
 \author[a]{Gustavo P. de Brito,}
 \author[a]{Astrid Eichhorn,}
 \author[a]{Rafael Robson Lino dos Santos}

 \affiliation[a]{CP3-Origins,  University  of  Southern  Denmark,  Campusvej  55,  DK-5230  Odense  M,  Denmark}

\emailAdd{gustavo@cp3.sdu.dk}
\emailAdd{eichhorn@cp3.sdu.dk}
\emailAdd{rado@cp3.sdu.dk}

\abstract{We discover a weak-gravity bound in scalar-gravity systems in the asymptotic-safety paradigm. The weak-gravity bound arises in these systems under the approximations we make, when gravitational fluctuations exceed a critical strength. Beyond this critical strength, gravitational fluctuations can generate complex fixed-point values in higher-order scalar interactions. Asymptotic safety can thus only be realized at sufficiently weak gravitational interactions. We find that within truncations of the matter-gravity dynamics, the fixed point lies beyond the critical strength, unless spinning matter, i.e., fermions and vectors, is also included in the model.
}

\maketitle

\section{Introduction}

The Standard Model of particle physics is very successful in describing a multitude of experimental results. However, from a theoretical point of view, its structure appears ad-hoc: Its symmetries, field content, charge assignments and interaction strengths are currently not explained. It is not known whether a fundamental principle exists, according to which the Standard Model is preferred over other  models with, e.g., different field content. A (partial) answer may come from embedding the Standard Model in a more fundamental theory, if such a fundamental theory has high predictive power. This idea  can be tested in distinct approaches to quantum gravity: \\
In string theory, there could be a huge number of matter models that can arise in the low-energy limit. It is not yet clear whether the Standard Model is actually one of them, see, e.g., \cite{Brennan:2017rbf,Palti:2019pca,vanBeest:2021lhn} for reviews of the swampland-program, 
and, e.g., \cite{Anderson:2011ns,Anderson:2012yf,Maharana:2012tu,Schellekens:2013bpa} for the construction of MSSM-like models from string theory. \\
\noindent In Loop Quantum Gravity, not enough about the interplay of quantum gravity with matter is known to say whether or not the framework has a high predictive (and thus constraining) power for matter models, see, e.g., \cite{Gambini:2004gu,Campiglia:2016bhw,Mansuroglu:2020dga}. \\
\noindent The situation is unique in asymptotically safe gravity-matter models, where strong indications for a high predictive power exist: the low-energy values of various couplings could either be bounded from above or be fixed precisely \cite{Shaposhnikov:2009pv,Harst:2011zx,Eichhorn:2017lry,Eichhorn:2017ylw,Eichhorn:2018whv}, an argument for the preference of four spacetime dimensions has been derived based on the gravity-matter interplay \cite{Eichhorn:2019yzm}, upper and lower bounds on the number of light fermions may exist \cite{Gies:2018jnv,Gies:2021upb,deBrito:2020dta} and certain discrete symmetries may be prohibited \cite{Ali:2020znq}. 

In this paper, we explore one specific question about the structure of the matter sector: could a gravity-matter model be asymptotically safe, if it contained just gravitational and scalar fields? We take two steps to develop a tentative answer. \\
In a first step, we build on previous work \cite{Narain:2009fy,Narain:2009gb,Zanusso:2009bs,Vacca:2010mj,Henz:2013oxa,Percacci:2015wwa,Labus:2015ska,Dona:2015tnf,Henz:2016aoh,Pawlowski:2018ixd,Wetterich:2019rsn,Eichhorn:2020sbo}  by following \cite{Eichhorn:2012va,Eichhorn:2016esv,Eichhorn:2017eht,Eichhorn:2017sok} and lifting an approximation that consists in neglecting shift-symmetric interactions.
This approximation makes a critical difference, because we find a fixed point that lies off the real axis in the complex plane. This result tentatively excludes gravity-scalar systems from being asymptotically safe. 
This conclusion holds under a set of assumptions and approximations that we spell out in more detail in this paper.\\
In a second step, we show that the mechanism that prevents gravity-scalar theories from being asymptotically safe may be circumvented by adding a sufficient number of fermion and/or vector fields to the theory. Together, these findings provide a tentative answer why viable matter sectors contain degrees of freedom beyond spin 0. 

To obtain this result, we rely on the interplay of two properties of asymptotically safe gravity matter models: first, it has been observed in \cite{Eichhorn:2016esv,Christiansen:2017gtg,Eichhorn:2017eht} that such models can exhibit a \emph{weak-gravity bound}. This bound is \emph{not} known to be related to the weak-gravity conjecture\footnote{See \cite{deAlwis:2019aud} for the asymptotically safe perspective on the weak-gravity conjecture.} \cite{ArkaniHamed:2006dz}. The bound arises because quantum gravity fluctuations can push matter systems beyond a point where matter couplings can become asymptotically safe. To avoid the bound, quantum gravity fluctuations must not be too strong; thus a fixed point for the gravitational couplings must not lie beyond the weak-gravity bound. 
This bound occurs in shift symmetric interactions, so far only considered in few works in gravity-scalar systems \cite{Eichhorn:2012va,Eichhorn:2013ug,Eichhorn:2016esv,Eichhorn:2017eht,Eichhorn:2017sok}. 

Second, it has been observed in \cite{Dona:2013qba,Labus:2015ska,Dona:2015tnf,Meibohm:2016mkp,Biemans:2017zca,Alkofer:2018fxj,Eichhorn:2018akn,Wetterich:2019zdo,Burger:2019upn} that the effective strength of quantum gravity fluctuations increases in response to an increasing number of scalar fields added to the model. In contrast, the effective strength of gravity fluctuations decreases in response to an increasing number of fermion and vector fields.

Taken together, these two observations could give rise to a mechanism by which asymptotic safety is prevented in gravity-matter systems beyond a critical number of scalar fields. This critical number is estimated in this paper for the first time. Additionally, we explore whether the matter content of the Standard Model results in asymptotic safety.\\

This paper is organized as follows: in Sec.~\ref{sec:ASQGmatt}, we explain the context of this work, which is the asymptotic safety framework for gravity and matter. In this framework, the weak-gravity bound limits the maximum strength of quantum gravity fluctuations. We discover that the weak-gravity bound exists in scalar-gravity systems and discuss it for a system with a single scalar field and gravity in Sec.~\ref{sec:WGB1scalar}. To extend the discussion from one to many scalars, in Sec.~\ref{sec:symmetries}, we first explain which symmetries constrain the interactions of many scalars coupled to gravity.  Based on these symmetry considerations, we analyze the weak-gravity bound in systems with more than one scalar. In Sec.~\ref{sec:upperbound}, we locate the fixed-point values for the gravitational  couplings  to establish bounds on the number of scalars. We conclude in Sec.~\ref{sec:conclusions}.  In App.~\ref{app:FRG} we  provide an introduction as well as technical details on the Functional Renormalization Group approach, exemplify our computational scheme with a shift and $\mathbb{Z}_2$ symmetric two-scalar field model, and establish our notation for the coupling with gravity. In App.~\ref{App:Pure_Matter}, we present our main results for the fixed-point structure in a two-scalar fields pure-matter setting. In App.~\ref{app:GaugeDep}, we analyze the gauge dependence of our results. Finally, in the App.~\ref{app:irrelevance}, we analyze the impact of quartic and shift symmetric  scalar self-interactions on  predictions of the Higgs quartic and Yukawa couplings  from asymptotic safety.

\section{Context of this work: Asymptotically safe gravity matter systems}\label{sec:ASQGmatt}

Asymptotic safety is a form of scale symmetry that is realized in the presence of quantum fluctuations. Generically,  quantum fluctuations break scale symmetry. It can be restored at special  values of the interaction strengths, where the effects of quantum fluctuations balance out. At these values, the Renormalization Group (RG) flow reaches an interacting fixed point.
Multiple examples for such interacting RG fixed points are known, see, e.g., \cite{Eichhorn:2018yfc} for a review. They include asymptotically safe theories in four dimensions which are under control in perturbation theory \cite{Litim:2014uca,Esbensen:2015cjw}. Further examples can be found in three-dimensional theories, see, e.g., \cite{Braun:2010tt,Eichhorn:2013zza,Dabelow:2019sty}. 
Based on the seminal work by Reuter \cite{Reuter:1996cp}, compelling evidence for asymptotic safety in gravity has been found, \cite{Souma:1999at,Lauscher:2001ya,Reuter:2001ag,Lauscher:2002sq,Litim:2003vp,Codello:2006in,Machado:2007ea,Codello:2008vh,Benedetti:2009rx,Eichhorn:2009ah,Manrique:2010am,Eichhorn:2010tb,Groh:2010ta,Dietz:2012ic,Christiansen:2012rx,Rechenberger:2012pm,Falls:2013bv,Ohta:2013uca,Eichhorn:2013xr,Falls:2014tra,Christiansen:2014raa,Demmel:2015oqa,Gies:2015tca,Christiansen:2015rva,Ohta:2015fcu,Ohta:2015efa,Falls:2015qga,Eichhorn:2015bna,Gies:2016con,Denz:2016qks,Biemans:2016rvp,Falls:2016msz,Falls:2016wsa,deAlwis:2017ysy,Christiansen:2017bsy,Falls:2017lst,Houthoff:2017oam,Falls:2017cze,Becker:2017tcx,Knorr:2017fus,Knorr:2017mhu,DeBrito:2018hur,Eichhorn:2018ydy,Falls:2018ylp,Bosma:2019aiu,Knorr:2019atm,Falls:2020qhj,Kluth:2020bdv,Knorr:2021slg,Bonanno:2021squ,Knorr:2021niv,Baldazzi:2021orb}, see \cite{Donoghue:2019clr,Bonanno:2020bil} for recent critical discussions of the state-of-the-art in the field and \cite{Eichhorn:2017egq,Percacci:2017fkn, Reuter:2019byg,Pereira:2019dbn,Reichert:2020mja,Pawlowski:2020qer} for reviews. Moreover, lattice techniques are being used to search for asymptotic safety, which manifests as a second-order phase transition \cite{Loll:2019rdj}. Additionally, tensor-model techniques enable a background-independent search for asymptotic safety \cite{Eichhorn:2018phj,Eichhorn:2019hsa}. 

Finding asymptotic safety in pure gravity does not suffice to build a phenomenologically viable quantum gravity theory. Such a theory must include matter, first because matter fields could spoil a pure gravitational fixed point, see, e.g.,  \cite{Dona:2013qba,Meibohm:2015twa,Dona:2015tnf,Biemans:2017zca,Alkofer:2018fxj,Wetterich:2019zdo}, and second because matter fields could lead to experimentally testable consequences of a fixed point, see \cite{Shaposhnikov:2009pv,Harst:2011zx,Eichhorn:2017ylw,Eichhorn:2017lry,Eichhorn:2018whv,Eichhorn:2020sbo}. Evidence for a fixed point in matter-gravity systems is mounting \cite{Eichhorn:2011pc,Dona:2013qba,Dona:2014pla,Meibohm:2015twa,Oda:2015sma,Dona:2015tnf,Wetterich:2016uxm,Eichhorn:2016vvy,Biemans:2017zca,Eichhorn:2017sok,Eichhorn:2017eht,Christiansen:2017cxa,Hamada:2017rvn,Pawlowski:2018ixd,Eichhorn:2018ydy,Alkofer:2018fxj,Eichhorn:2018akn,Eichhorn:2018nda,Wetterich:2019zdo,deBrito:2019epw,deBrito:2019umw,Eichhorn:2020sbo,Burger:2019upn,Daas:2021abx}, with intriguing phenomenological consequences \cite{Shaposhnikov:2009pv,Harst:2011zx,Eichhorn:2017lry,Eichhorn:2017ylw,Eichhorn:2017als,Eichhorn:2018whv,Reichert:2019car,Alkofer:2020vtb,Eichhorn:2020kca}. More recently,  lattice studies also include  gravity-matter systems \cite{Jha:2018xjh,Catterall:2018dns,Dai:2021fqb,Ambjorn:2021fkp,Ambjorn:2021uge}. 

In this work, we employ the functional Renormalization Group (RG) technique \cite{Wetterich:1992yh,Morris:1993qb,Reuter:1996cp}, see \cite{Dupuis:2020fhh} for a recent review and App.~\ref{app:FRG} for further technical details. With this technique we derive beta functions that encode the scale-dependent change of couplings. The scale-dependence is caused by quantum fluctuations that are accounted for in a scale-dependent fashion. 

Specifically, the beta functions have a two-fold use: first, searching for points in the space of couplings where all beta functions vanish provides the coordinates of a fixed point, i.e., the interaction strengths at which asymptotic safety is realized. Second, using appropriate initial conditions, the scale-dependence of the couplings can be calculated from the beta functions. In particular, one can construct RG trajectories which feature i) a scale-symmetric regime in the ultraviolet (UV), ii) a transition scale at which couplings depart from scale-symmetry along the relevant (infrared (IR) repulsive) directions of a fixed point, iii) an  IR regime in which the values of irrelevant (IR attractive) couplings are determined by their origin in the asymptotically safe regime. For gravity-matter systems, the transition scale typically agrees with the Planck scale, at which quantum gravity fluctuations decouple dynamically. 

In this paper, we focus on the first regime to explore under which conditions an asymptotically safe fixed point can exist in gravity-matter systems.

Without providing extensive details on technical questions in the main text (see App.~\ref{app:FRG} for that), let us mention two main sources of systematic uncertainties in our results: \\
First, we do not account for the Lorentzian nature of quantum gravity fluctuations, since the functional RG is based on an infrared cutoff that cannot straightforwardly be imposed in Lorentzian signature, see \cite{Manrique:2011jc} for a first study. Under appropriate analyticity assumptions \cite{Draper:2020bop,Platania:2020knd}, a continuation of Euclidean results to Lorentzian signature could be viable \cite{Bonanno:2021squ}. \\
Second, we truncate the space of couplings to a finite dimensional subspace of the infinite-dimensional space that encompasses all possible dynamics in a Wilsonian approach to quantum field theory\footnote{A related source of systematic uncertainties is renormalization scheme dependence. Universal quantities and statements, such as the existence of a fixed point and critical exponents, are not expected to depend on the choice of scheme. This is always true for untruncated theory spaces. However, truncations usually lead to dependence on unphysical choices, such as that of the scheme, and consequently scheme dependence is another source of systematic uncertainties in our results. Specifically, in the present case scheme dependence refers to the choice of regulator shape function.}. Our truncation is based on a bootstrap argument \cite{Falls:2013bv} for which indications have been found in gravity-matter systems, e.g., in \cite{Eichhorn:2016vvy,Eichhorn:2017sok,Eichhorn:2018ydy,Eichhorn:2018nda}: if an asymptotically safe fixed point has near-canonical scaling behavior, then canonically higher-order couplings are subleading compared to the canonically most relevant ones. 
In the context of our work, it is important to differentiate between interactions with different symmetry structures: in each sector determined by a given symmetry, the leading-order couplings in that sector should be accounted for. We focus on scalars with a $\mathbb{Z}_2$ reflection symmetry and shift symmetry.
For this case, the leading order operators are dimension-8 operators. In a more naive application of the bootstrap strategy which does not account for symmetries carefully, these might be neglected in a truncation. As we will show, they may actually be critical for a comprehensive understanding of gravity-scalar systems.

\section{Weak-gravity bound for one scalar}\label{sec:WGB1scalar}

In this section, we consider a single real scalar field $\phi$, minimally coupled to quantum gravity, with a shift-symmetric quartic self-interaction. Our truncation of the dynamics in four dimensions is given by
\begin{eqnarray}
\Gamma_k \!=\! \frac{Z_{\phi}}{2}\int_x\!\!\sqrt{g}g^{\mu\nu}\partial_{\mu}\phi\partial_{\nu}\phi + \frac{\bar{g}}{8}\int_x\!\!\sqrt{g}g^{\mu\nu}g^{\kappa\lambda}\partial_{\mu}\phi \partial_{\nu}\phi \partial_{\kappa}\phi\partial_{\lambda}\phi - \frac{1}{16\pi G_N} \int_x\!\!\sqrt{g}\left(R- 2\bar{\Lambda} \right) + \Gamma_{\rm gf}. \!\!\nonumber\\
\label{eq:truncationsinglescalar}
\end{eqnarray}
The restriction to shift symmetry will be justified in detail in Sec.~\ref{sec:symmetries} below. In short, shift symmetry is respected by quantum gravitational fluctuations \cite{Eichhorn:2012va,Eichhorn:2013ug,Eichhorn:2016esv,Eichhorn:2017eht,Eichhorn:2017sok}. Thus, a shift-symmetry breaking scalar potential is not generated by quantum gravity \cite{Narain:2009fy,Labus:2015ska,Eichhorn:2017als,Wetterich:2019rsn,Eichhorn:2020sbo}. In contrast, shift-symmetric, momentum-dependent interactions are generated.

The absence of non-minimal, shift symmetric couplings, such as that investigated in \cite{Eichhorn:2017sok} as well as higher-order shift symmetric interactions is a choice of truncation. We work under the hypothesis that the subleading impact of non-minimal shift-symmetric interactions observed in \cite{Eichhorn:2017sok} extends to our case, in which shift-symmetric self-interactions are included.

In Eq.~\eqref{eq:truncationsinglescalar}, an Einstein-Hilbert term is included for the gravitational dynamics together with a gauge-fixing term. We work in Landau-DeWitt gauge. 

In this setting, the beta function for $g = \bar{g}\,k^{4}/Z_\phi^2$ is given by
\begin{eqnarray}\label{eq:betagsinglescalar}
	\beta_g &=& 4 g  + \frac{9}{64\pi^2} g^2 - \left( \frac{10}{3 \pi  (1-2 \Lambda)^2} - \frac{5}{18 \pi  (1-4 \Lambda/3)^2} - \frac{4}{9 \pi  (1-4 \Lambda/3)} \right) g  \,G  \nonumber \\
	&+& \left( \frac{640}{9 (1-2 \Lambda)^3}+\frac{4}{9 \left(1-4 \Lambda /3\right)^3}+\frac{4}{9 \left(1-4 \Lambda /3 \right)^2} \right) G^2  \,,
\end{eqnarray} 
Here, we have made the transition to dimensionless gravitational couplings
\begin{eqnarray}
	G= G_N\, k^2 \qquad \textmd{and} \qquad \Lambda = \bar{\Lambda}\, k^{-2}.
\end{eqnarray}
Their fixed-point values $G_{\ast}$ and $\Lambda_{\ast}$ are treated as free parameters throughout large parts of this paper.
The solutions for the fixed point in $g$  are
\begin{eqnarray}
	\hspace*{-.4cm} g_{\ast, \pm} \!= -\frac{32 \pi^2}{9} \bigg[ 4 \!-\!  \frac{10 \,G_*}{3 \pi  (1-2 \Lambda_*)^2} \!+\!  \frac{5 \,G_*}{18 \pi  (1-4 \Lambda_*/3)^2} \!+\!  \frac{4 \,G_*}{9 \pi  (1-4 \Lambda_*/3)}   \pm
	\Delta(G_*,\Lambda_*)^{1/2} \bigg] ,\,
\end{eqnarray}
where
\begin{eqnarray}
	\Delta(G_*,\Lambda_*) &=& -\frac{G^2_*}{4 \pi ^2} \left(\frac{160}{(1-2 \Lambda_* )^3}+\frac{9}{(3-4 \Lambda_* )^2}+\frac{27}{(3-4 \Lambda_* )^3}\right) \nonumber \\
	&+&\bigg( 4 \!-\!  \frac{10 \,G_*}{3 \pi  (1-2 \Lambda_*)^2} \!+\!  \frac{5 \,G_*}{18 \pi  (1-4 \Lambda_*/3)^2} \!+\!  \frac{4 \,G_*}{9 \pi  (1-4 \Lambda_*/3)} \bigg)^2 \, .
\end{eqnarray}
In the presence of nonzero gravitational fixed-point values, $g_{\ast}$ is nonzero  for both solutions  and $g_{\ast, -}$ reduces to the free fixed point for $G_{\ast} \rightarrow 0$. We refer to such interactions as gravity-induced interactions. They are the \emph{unavoidable} matter-self interactions when gravity is present.

\begin{figure}[!t]
\centering
\hspace*{-.7cm}\includegraphics[height=6.5cm]{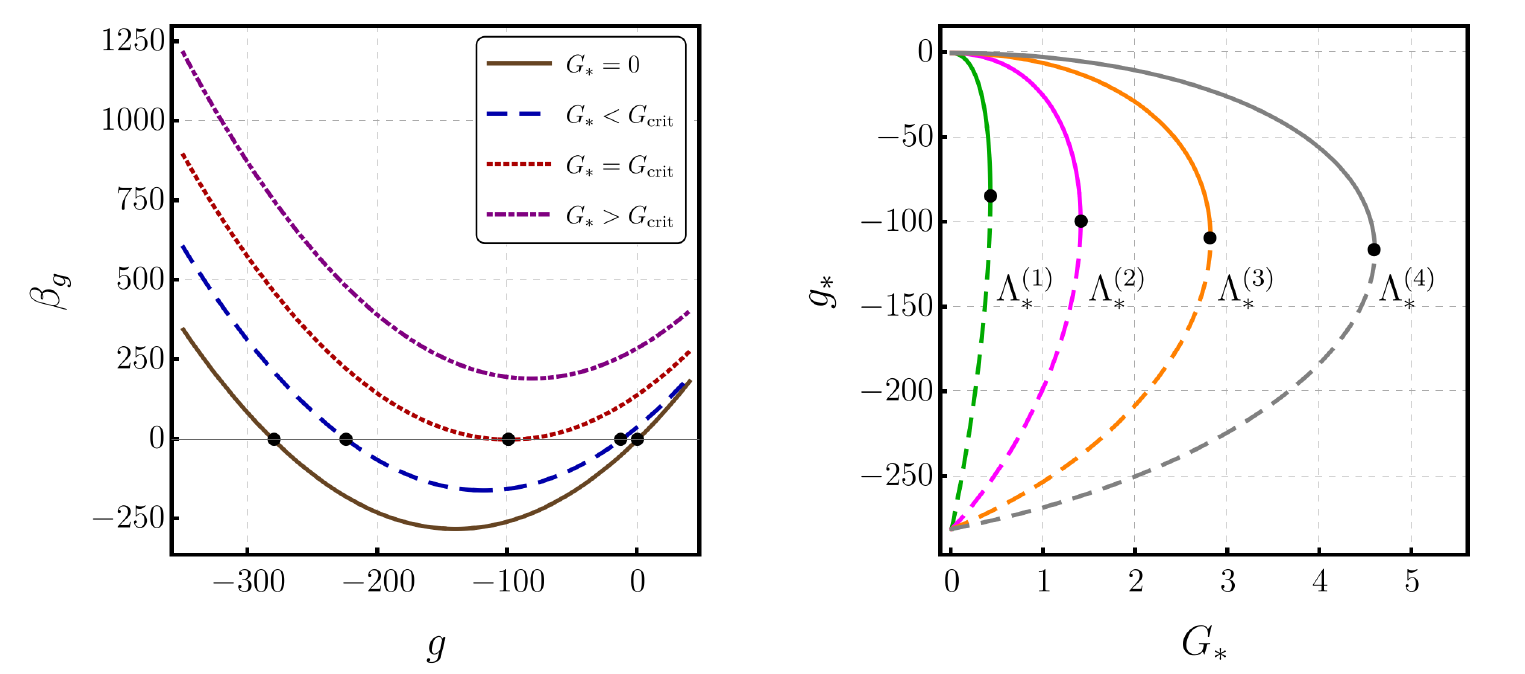}
\caption{\label{fig:FPcollision}  Left panel: We plot the beta function as a function of $g$ for $\Lambda_*=0$ and different values of $G_*$. 
When $G_*=0$ (brown continuous line), there are two real zeros of $\beta_g$. By increasing $G_*$, the distance between  the fixed points decreases, see, e.g., $G_{\ast}<G_{\textmd{crit}}$ (blue dashed line). These fixed points collide when $G_*= G_{\textmd{crit}}$ (red dotted line). For $G_*> G_{\textmd{crit}}$ (purple dot-dashed line), the solutions for $\beta_g = 0$ are complex. Right panel: We show the collision of the shifted Gaussian fixed point (continuous lines) and the non-Gaussian fixed point (dashed lines) for different values of $\Lambda_{*}$: from left to right, $\Lambda^{(1)}_{*}=0.25$ (green), $\Lambda^{(2)}_{*}=0$ (magenta),  $\Lambda^{(3)}_{*}=-0.25$ (orange), and  $\Lambda^{(4)}_{*}=-0.5$ (gray). 
}
\end{figure}

\begin{figure}[!t]
\centering
\hspace*{-.75cm}\includegraphics[height=6.0cm]{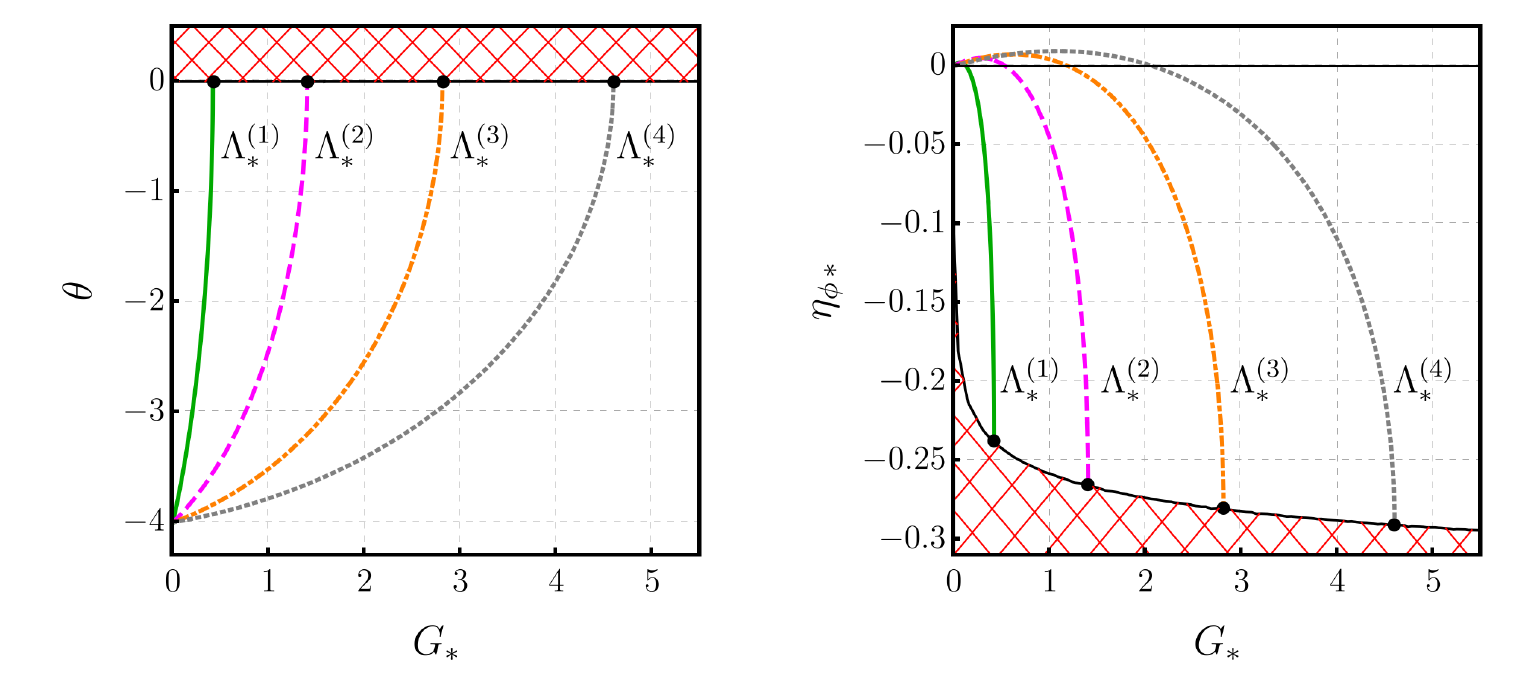}
\caption{\label{fig:fulleta}  Left panel: We show the critical exponent $\theta$ at the shifted Gaussian fixed point for different values of $\Lambda_{\ast}$: $\Lambda^{(1)}_{*}=0.25$ (green continuous line), $\Lambda^{(2)}_{*}=0$ (magenta  dashed line),  $\Lambda^{(3)}_{*}=-0.25$ (orange  dot-dashed line), and  $\Lambda^{(4)}_{*}=-0.5$ (gray dotted line). $\theta$ tends to zero when $G_{\ast}\rightarrow G_{\rm crit}$, as expected at a fixed-point collision. Right panel: We show  the anomalous dimension $\eta_{\phi}$ computed at the shifted Gaussian fixed point for the same values of $\Lambda_{*}$.  For large enough $G_{\ast}$, $\eta_{\phi\, \ast}$ can assume negative values due to loop diagrams including the new matter self-interaction. In App.~\ref{app:irrelevance}, we discuss how this impacts the irrelevance of the Higgs quartic and the Yukawa coupling in asymptotically safe gravity-matter models. }
\end{figure}

The dynamical mechanism underlying the weak-gravity bound can be seen at $\Lambda_{\ast}=0$ and then generalized to $\Lambda_{\ast} \neq 0$. 
At $\Lambda_{\ast}=0$, the beta-function Eq.~\eqref{eq:betagsinglescalar} is a parabola in $g$ with positive coefficients in the quadratic and the zeroth-order term, cf.~Fig.~\ref{fig:FPcollision}. At $G_{\ast}=0$, the zeroth order term vanishes, guaranteeing the existence of two real zeros of $\beta_g$, one of which lies at $g_{\ast}=0$. At $G_{\ast}>0$, the zeroth-order term is positive, thus shifting the Gaussian fixed point at $g_{\ast}=0$ to a \emph{shifted Gaussian fixed point} at $g_{\ast} \neq 0$. At $G_{\rm crit} \approx 1.4$, this fixed point collides with the second interacting fixed point, see Fig.~\ref{fig:FPcollision}. 
At this collision point, $\beta_g$ must have a degenerate zero, such that the 
 critical exponent, $\theta = - \frac{\partial \beta_g}{\partial g}\Big|_{g=g_{\ast}}$, vanishes, cf.~Fig.~\ref{fig:fulleta}.
Beyond $G_{\rm crit}$, both solutions of $\beta_g=0$ are complex. 

We require a real fixed-point value in all couplings in the theory. Accordingly, within this truncation, a viable fixed-point value for $G$ must lie at $G_{\ast}< G_{\rm crit}$. The strength of gravitational fluctuations, which are proportional to $G_{\ast}$, is therefore bounded from above, i.e., asymptotically safe gravity must satisfy a \emph{weak-gravity bound}.

\begin{figure}[!t]
	\centering
	\hspace*{-.75cm}\includegraphics[height=7.0cm]{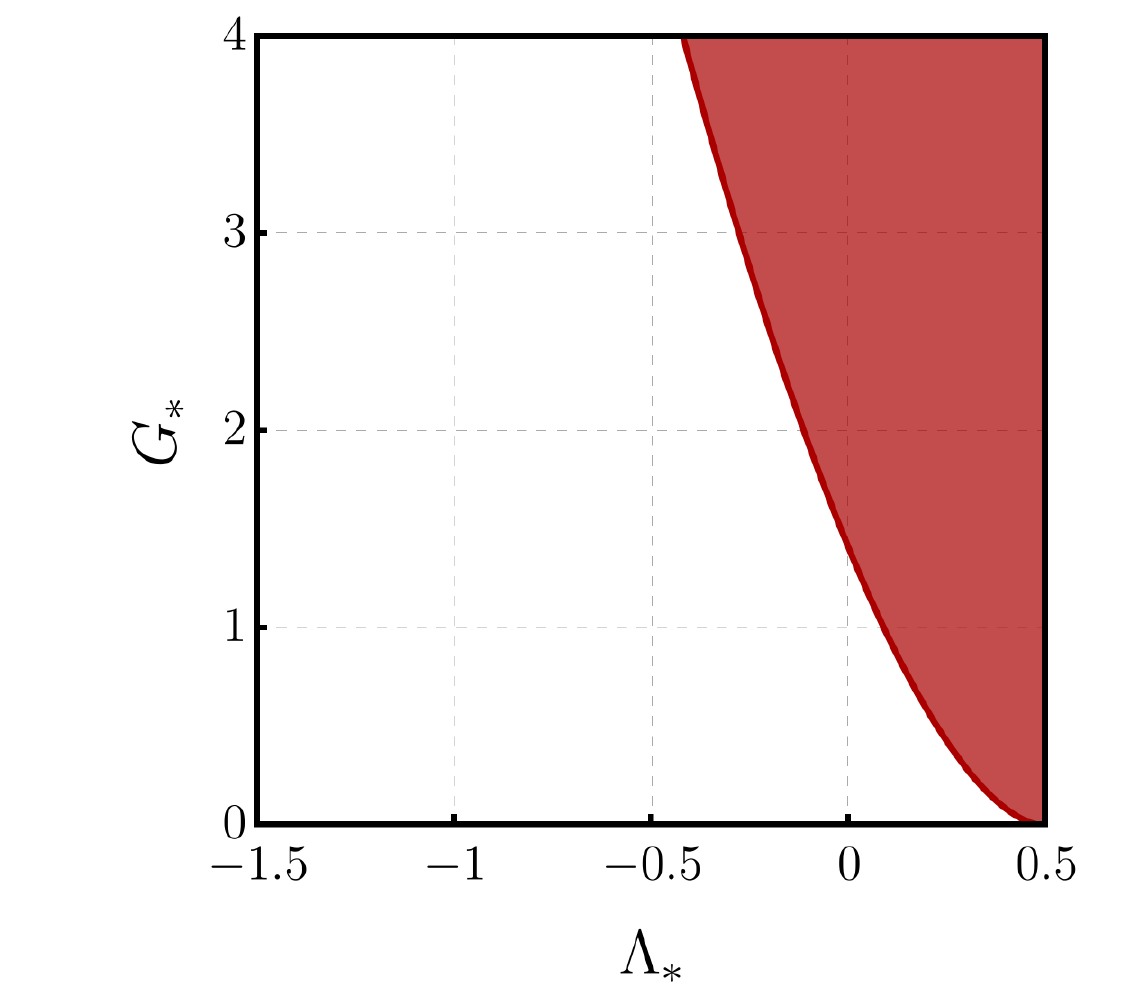}
	\caption{\label{fig:WGB_1Scalar} In red (white), the region forbidden (allowed) by the weak-gravity bound. A viable fixed-point value for $G$ must lie at $G_{\ast}< G_{\rm crit}(\Lambda_{\ast})$.}
\end{figure}

At $\Lambda_{\ast} \neq 0$, the coefficients of the zeroth-order and the linear term in $g$ change. As the zeroth-order term remains positive for $\Lambda_{\ast} \in (-\infty,1/2)$, which is the relevant range for fixed-point values. The weak-gravity bound persists, with $G_{\rm crit} = G_{\rm crit}(\Lambda_{\ast})$. Explicitly, the boundary separating the excluded strong-gravity from the allowed weak-gravity regime lies at
\begin{eqnarray}
	G_{\rm crit}(\Lambda_{\ast})= \frac{8\pi}{ \frac{20/3}{(1-2 \Lambda_{\ast} )^2} - \frac{8/9}{(1-4 \Lambda_{\ast}/3)} -\frac{5/9}{(1-4 \Lambda_{\ast}/3)^2} + \sqrt{\frac{160}{(1-2 \Lambda_{\ast})^3}+\frac{1}{(1-4 \Lambda_{\ast}/3)^2}+\frac{3}{(1-4 \Lambda_{\ast}/3)^3}}}.
	\quad\label{eq:GcritLambda}
\end{eqnarray}
As $\Lambda_{\ast}$ decreases, $G_{\rm crit}$ increases, because a combination of $G_{\ast}$ and $\Lambda_{\ast}$ enters the beta functions. The effective strength of metric fluctuations increases with increasing $G_{\ast}$ (at fixed $\Lambda_{\ast}$) and decreases with decreasing $\Lambda_{\ast}$ (at fixed $G_{\ast}$), because $\Lambda_{\ast}$ enters the denominator of the graviton propagator.  Negative values of $\Lambda_{\ast}$ act like a mass-like suppression. 
The effective strength of gravity is constant along the curve in Eq.~\eqref{eq:GcritLambda} for which $G_{\rm crit}$ increases when $\Lambda_{\ast}$ decreases, see Fig.~\ref{fig:WGB_1Scalar}.

Our choice of truncation affects the location of the weak-gravity bound: At finite values of higher-order gravitational as well as matter couplings, the critical curve $G_{\rm crit} (\Lambda_{\ast})$ will be deformed. We can estimate the robustness of our result for $G_{\rm crit} (\Lambda_{\ast})$ by exploring the gauge dependence. We find mild variations which might be interpreted as a hint that our result for $G_{\rm crit} (\Lambda_{\ast})$ is rather robust, cf.~App.~\ref{app:GaugeDep}.

\section{From one to many scalars }\label{sec:symmetries}

The weak-gravity bound exists in systems with different field content.
For instance, it occurs in systems containing spin-1/2 fermions coupled to scalars \cite{Eichhorn:2016esv, Eichhorn:2017eht} and systems with vectors \cite{Christiansen:2017gtg,Eichhorn:2019yzm}. 
In the presence of several matter fields, multiple gravity-induced interaction channel may exist, which may be coupled to each other and thereby trigger or prevent the weak-gravity bound, as in \cite{Eichhorn:2016esv, Eichhorn:2017eht}. The channels that are relevant in this context are those whose couplings cannot be set to zero in the presence of gravity. In \cite{Eichhorn:2017eht}, it was hypothesized that these are the channels determined by the maximum symmetry of the kinetic term. 
Conversely, any coupling that breaks this symmetry can be consistently set to zero, unless one chooses a regulator that breaks these symmetries. This is \emph{not} a choice of truncation, it is a choice that the matter-gravity dynamics makes for its fixed-point structure and entire RG flow. 

In order to generalize the weak-gravity bound from one to many scalars, we  must first understand the symmetries which select the interactions that give rise to the weak-gravity bound.

\subsection{Symmetry considerations}

Here, we will analyze induced channels in systems with $N_{\phi}$ real scalars to test the hypothesis that induced interactions respect the maximum symmetry of the kinetic term.
For our field content, the maximum symmetry of the kinetic term consists of:
\begin{itemize}
	\item $N_{\phi}$ independent $\mathbb{Z}_2$ symmetries, where  $\phi_i \rightarrow -\phi_i$ for the $i^{th}$ independent $\mathbb{Z}_2$ symmetry,
	\item $N_{\phi}$ independent shift symmetries, where $\phi_i \rightarrow \phi_i + a_i$ for the $i^{th}$ independent shift symmetry,
	\item O$(N_{\phi})$ symmetry, where $\phi_i$ transforms as a vector in the fundamental representation.
\end{itemize}

Couplings that satisfy these symmetries can be set to zero in the absence of gravity, i.e., they feature a Gaussian fixed point that is non-interacting\footnote{Additionally, they may exhibit interacting fixed points, see App.~\ref{App:Pure_Matter}.}. They 
cannot be set to zero in the presence of gravity \cite{Eichhorn:2012va, Eichhorn:2017eht}.
Quantum gravity fluctuations shift the
Gaussian fixed point
to a shifted Gaussian fixed point (sGFP) that is interacting. For large enough shifts, the sGFP is not guaranteed to be real, resulting in a weak-gravity bound.

In a theory space with less symmetry, the fixed-point values are divided into two classes: maximally symmetric interactions are nonzero and maximum-symmetry breaking interactions are vanishing.
Explicit confirmations of this division were found in the literature: 
first, a shift-symmetry breaking scalar potential and a non-minimal coupling $R\,\phi^2$ vanish at the fixed point with gravity\footnote{In the presence of fermions and gauge fields, it is possible to realize a non-zero potential and nonminimal coupling by a \emph{choice of fixed point} outside the maximally symmetric theory space, see, e.g., \cite{Eichhorn:2017ylw,Eichhorn:2019dhg,Wetterich:2019rsn,Eichhorn:2020sbo}. Distinct universality classes with distinct symmetries may thus exist for gravity-matter systems.} \cite{Narain:2009fy,Narain:2009gb,Percacci:2015wwa,Labus:2015ska,Oda:2015sma,Eichhorn:2017als,Eichhorn:2020sbo}. 
Second, the maximally symmetric interactions $R_{\mu\nu}\partial^{\mu}\phi\partial^{\nu}\phi$ 
and $g^{\mu\nu}\partial_{\mu}\phi\partial_{\nu}\phi\,g^{\kappa\lambda}\partial_{\kappa}\phi\partial_{\lambda}\phi$
do not vanish at the fixed point with gravity \cite{Eichhorn:2017sok,Eichhorn:2012va}. 
 
The maximally symmetric theory space is closed under the RG flow, even in the presence of quantum gravity\footnote{The realization of global symmetries in the presence of gravity in the asymptotic-safety framework might not persist in Lorentzian signature, or might be a property that sets asymptotically safe gravity apart from other approaches to quantum gravity \cite{Banks:1988yz,Kamionkowski:1992mf,Kallosh:1995hi}, see also \cite{Eichhorn:2020sbo} for a discussion of this question.}. In the asymptotic safety framework, quantum gravity does not induce maximum-symmetry breaking couplings which can thus consistently be set to zero. In the larger theory space with less symmetry, maximum-symmetry breaking couplings may also have non-vanishing fixed-point values. Different choices of symmetries may be compatible with the asymptotic-safety requirement and several gravity-matter universality classes with different symmetries may exist. All such universality classes contain at least the interactions we explore here (and potentially more), thus our study is relevant to any scalar-gravity system.

To explicitly confirm that the symmetries listed above are indeed respected in maximally symmetric matter-gravity systems, we focus on the case $N_{\phi}=2$. 

As a first check, we confirm the $\mathbb{Z}_2$ symmetries. They are realized if all uneven orders in the field expansion of the flow equation vanish. We find explicitly that this is the case. This follows directly from the fact that all vertices are quadratic in the scalar field and that there are not mixed propagators between any of the fields.

As a second check, we confirm the shift symmetries. They are realized if the $n$th order in the field expansion of the flow equation starts at $n$th order in the derivative expansion and each interaction can be written as a contraction of $\partial_{\mu}\phi_i$ (upon partial integrations). 
We focus on the first two orders in the field expansion, namely quadratic and quartic scalar interactions. For the quadratic interaction, the second order of the derivative expansion (at two derivatives) is the first non-vanishing order. For the quartic interaction, the fourth order in the derivative expansion (at four derivatives) is the first non-vanishing order. In both cases, each external scalar field appears together with its momentum, thus the projection of the flow equation outside the shift-symmetric theory space vanishes. 
Beyond these two orders in the field that we have checked explicitly, shift symmetry must also hold. This follows directly, as the vertices that cannot be set to zero are derived from the kinetic term, where each scalar field always appears together with its momentum.

\begin{figure}[!t]
	\centering
	\hspace*{-.75cm}\includegraphics[height=7.0cm]{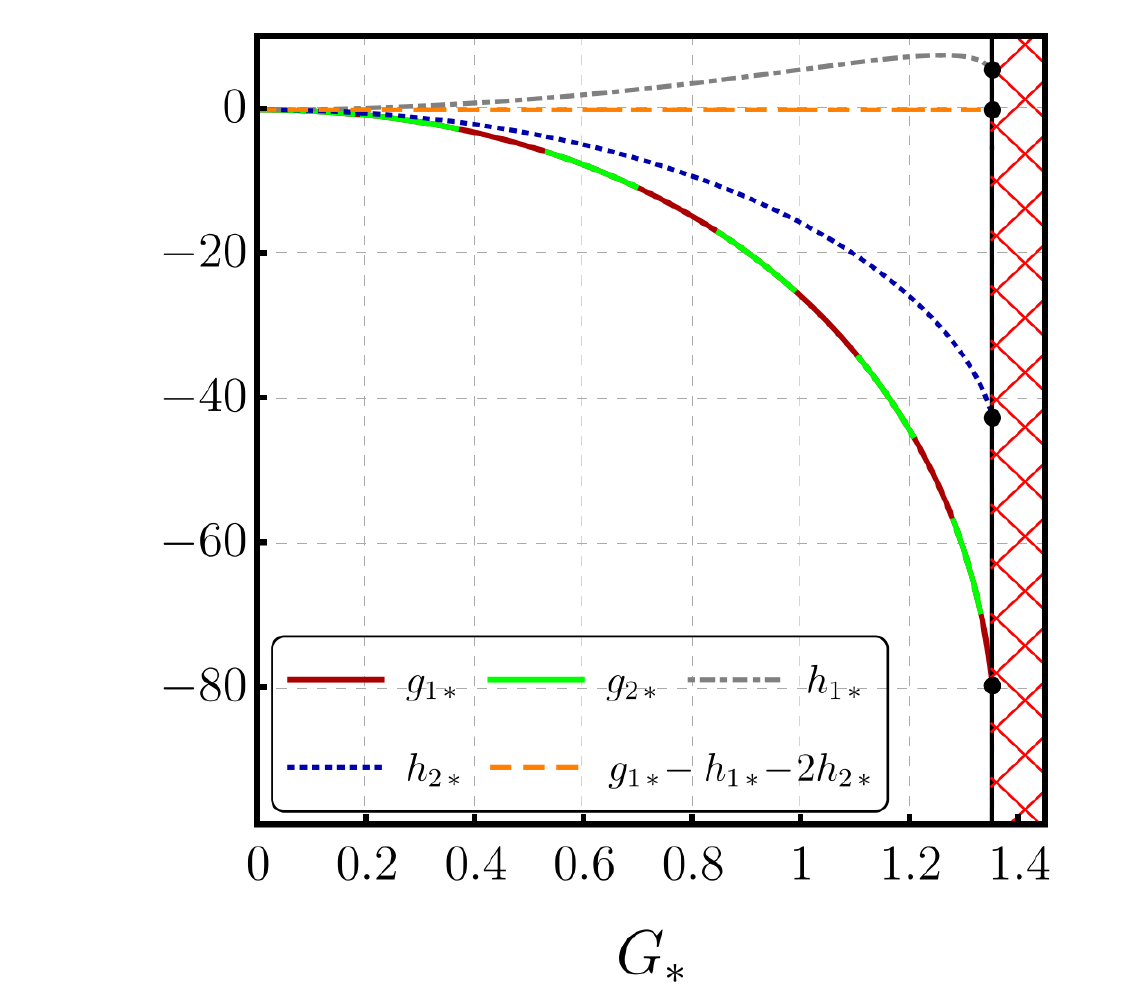}\quad
	\caption{\label{fig:O2FP} We show the shifted Gaussian fixed point for the four couplings as functions of $G_*$ for $N_\phi = 2 $ and $\Lambda_*=0$. In the presence of gravity, the Gaussian fixed point is shifted, but the couplings still respect the O(2) symmetry relations $g_{1\,*} - g_{2\,*} = 0$ and $g_{1\,*} - h_{1\,*} - 2 h_{2\,*} = 0$, for $G_*<G_{\rm crit}$.}
\end{figure}

As a third check, we confirm the O(2) symmetry. It is realized if relations hold between the shift-symmetric interactions of the two scalars. These relations reduce the number of independent couplings in
the most general dynamics invariant under shift- and $\mathbb{Z}_2$-symmetry. We truncate this dynamics to include only scalar self-interactions at fourth order in the field and derivative expansion. These are the leading interactions according to canonical power counting.
\begin{align}\label{Truncation_2scalar}
	\Gamma_k^{\textmd{2-Scalars}} =  \int_x \sqrt{g} \,
	&\left( \frac{Z_{1}}{2} g^{\mu\nu} \pt_\mu \phi_1 \pt_\nu \phi_1 +
	\frac{Z_{2}}{2} g^{\mu\nu} \pt_\mu \phi_2 \pt_\nu \phi_2  \right)  \\
	+ \int_x \sqrt{g} \,
	&\bigg( \frac{\bar{g}_{1}}{8} g^{\mu\nu} g^{\alpha\beta} \pt_\mu \phi_1 \pt_\nu \phi_1  \pt_\alpha \phi_1 \pt_\beta \phi_1 + \frac{\bar{g}_{2}}{8} g^{\mu\nu} g^{\alpha\beta} \pt_\mu \phi_2 \pt_\nu \phi_2  \pt_\alpha \phi_2 \pt_\beta \phi_2 
	\nonumber\\ 
	& + \frac{\bar{h}_{1}}{4} g^{\mu\nu}g^{\alpha\beta}\pt_\mu\phi_1\pt_\nu\phi_1\pt_\alpha\phi_2\pt_\beta\phi_2 + \frac{\bar{h}_{2}}{2} g^{\mu\nu}g^{\alpha\beta} \pt_\mu\phi_1\pt_\nu\phi_2\pt_\alpha\phi_1 \pt_\beta\phi_2  \bigg) . \nonumber
\end{align}
For generic values of the couplings, there is no O(2)-symmetry. The O(2)-symmetry  requires the following relations
\begin{align}\label{eq:O(2)-relations}
	Z_{1} - Z_{2} = 0 \,, \quad g_1 - g_2 = 0  \quad \textmd{and} \quad g_1 - h_1 - 2 h_2 = 0  \,,
\end{align}
for the dimensionless couplings $g_i =k^4 \bar{g}_i/Z_i^2$ and $h_i = k^4 \bar{h}_i/(Z_1\,Z_2)$. Based on our symmetry considerations, we expect that under the impact of quantum gravity, a fixed point satisfies these relations \eqref{eq:O(2)-relations}. This can be checked by inspecting the beta functions
\begin{eqnarray}
	\hspace*{-.5cm}\beta_{g_1}\!& =&\!(4+2\eta_1)g_1 \!+\! \frac{5g_1^2 + 2h_1^2 + h_2^2+2h_1h_2}{64\pi^2} - \frac{17}{6\pi} g_1\,G + 72 \,G^2 \,,\label{eq:betag1}\\
	\hspace*{-.5cm}\beta_{g_2}\!& =&\! (4+2\eta_2)g_2 \!+ \frac{5g_2^2 + 2h_1^2 + h_2^2+2h_1h_2}{64\pi^2} - \frac{17}{6\pi} g_2\,G + 72 \,G^2 \,,\\
	\hspace*{-.5cm}\beta_{h_1}\!&=&\! (4+\eta_1+\eta_2) h_1 \!+\! \frac{(g_1+g_2)(9h_1+4h_2)\!+\!(h_1+h_2)^2}{192\pi^2} \!-\! \frac{17(2h_1+h_2)}{18 \pi}G \!-\! \frac{632}{27} G^2 \,, \\
	\hspace*{-.5cm}\beta_{h_2}\!&=&\!(4+\eta_1+\eta_2) h_2 \!+\! \frac{h_1^2+8h_1h_2 + 19 h_2^2 \!+\! h_2(g_1+g_2)}{192\pi^2} \!-\! \frac{17(h_1+5h_2)}{36 \pi}G \!-\! \frac{1288}{27} G^2 \,,\label{eq:betah2}
\end{eqnarray}
with anomalous dimensions
\begin{eqnarray}
	\eta_1 = \frac{3g_1 + 2h_1+h_2}{96\pi^2} + \frac{G}{9\pi}
	\qquad \textmd{and} \qquad 
	\eta_2 = \frac{3g_2 + 2h_1+h_2}{96\pi^2} + \frac{G}{9\pi} \,.\label{eq:anomdim2}
\end{eqnarray}
 For simplicity, we  only  show RG-equations evaluated at $\Lambda = 0$. The results  concerning the symmetry structure  hold at non-vanishing $\Lambda$.
In the absence of gravity, at $G=0$, the system features a Gaussian fixed point, $g_{1,2\,\ast}=0=h_{1,2\,\ast}$ that satisfies the O(2)-symmetry relations \eqref{eq:O(2)-relations} trivially. In the presence of gravity, at $G>0$, the Gaussian fixed point gets shifted to an interacting fixed point that still satisfies the O(2) symmetry relations, cf.~Fig.~\ref{fig:O2FP}.

More generally, RG-consistency requires the symmetry relations \eqref{eq:O(2)-relations} along the flow, meaning that the corresponding flow equations must satisfy 
\begin{eqnarray}
	\hspace{-1cm} (\eta_{1} - \eta_{2})|_{\mathcal{P}_{\textmd{O}(2)}} &=& 0 \,,\label{eq:RG-consistency1}\\
	\hspace{-1cm} (\beta_{g_1} - \beta_{g_2} ) |_{\mathcal{P}_{\textmd{O}(2)}} & =&0 \,, \label{eq:RG-consistency2}\\
	\hspace{-1cm} (\beta_{g_1} - \beta_{h_1} - 2\beta_{h_2}) |_{\mathcal{P}_{\textmd{O}(2)}} & =&0 \,. \label{eq:RG-consistency3}
\end{eqnarray}
where $\mathcal{P}_{\textmd{O}(2)}$ indicates a projection onto the O(2) symmetric subspace. The explicit beta functions in Eq.~\eqref{eq:betag1}-\eqref{eq:betah2} and Eq.~\eqref{eq:anomdim2} satisfy these relations. Explicitly, it holds that
\begin{eqnarray}
	\hspace{-1cm} \eta_1 - \eta_2 &=& \frac{g_1 - g_2}{32\pi^2},\\
	\hspace{-1cm} \beta_{g_1} - \beta_{g_2} &= &2\,(\eta_1g_1-\eta_2g_2) + \left( 4 - \frac{17\, G}{6\pi} + \frac{5}{64\pi^2}(g_1 + g_2) \right) (g_1 - g_2)\,, \\
	\hspace{-1cm} \beta_{g_1} - \beta_{h_1} - 2\beta_{h_2} &= &g_1 (\eta_1 - \eta_2) + \frac{1}{64\pi^2} (3 h_1 + 2 h_2) (g_1 - g_2) \nonumber \\
	&{}&+\left( 4+ \eta_1 + \eta_2  - \frac{17\,G}{6\pi} + \frac{5g_1 - h_1 + 6 h_2}{64 \pi^2} \right) (g_1 - h_1 - 2\,h_2),
\end{eqnarray}
which immediately reduces to Eq.~\eqref{eq:RG-consistency1}-\eqref{eq:RG-consistency3} once we project the right-hand side onto the subspace defined by  \eqref{eq:O(2)-relations}.

Our explicit analysis confirms that the shifted Gaussian fixed point satisfies the three symmetries listed above. This limits the possible interaction channels that have to be taken into account when exploring the weak-gravity bound, since it is a bound that occurs already within the maximally symmetric theory space, i.e., with as many interactions as possible set to zero (and thus safe from the mechanism underlying the weak-gravity bound).

\subsection{Weak-gravity bound for many scalars}

Motivated by the discussion of the previous section, we now consider a shift-, $\mathbb{Z}_2-$ and O($N_{\phi}$) symmetric model involving $N_\phi$ scalar fields coupled to gravity, with the gravity dynamics given by a gauge-fixed Einstein-Hilbert action and the matter dynamics given by
\begin{align}\label{eq:Truncation_O(N)}
	\Gamma_k^{\textmd{O}(N_\phi)-\textmd{Matter}} &\!=\! \int_x \!\sqrt{g} \left( \frac{Z_{\phi}}{2} g^{\mu\nu} \partial_\mu \Phi^T \partial_\nu \Phi  \right)  \\ 
	&\!\!+\int_x \!\sqrt{g} \left(  
	\frac{\bar{\mathfrak{g}}_{1} }{8 N_\phi} g^{\mu\nu} g^{\alpha\beta}
	\partial_\mu \Phi^T \partial_\nu \Phi \,\partial_\alpha \Phi^T \partial_\beta \Phi +
	\frac{\bar{\mathfrak{g}}_{2} }{4 N_\phi} g^{\mu\nu} g^{\alpha\beta} 
	\partial_\mu \Phi^T \partial_\alpha \Phi \,\partial_\nu \Phi^T \partial_\beta \Phi \right) \!. \nonumber
\end{align}  
$\Phi = \left(\phi_1 \, \cdots \, \phi_N \right)^T$ is an O$(N_{\phi})$-multiplet. 
Based on this truncation for $\Gamma_k$ we arrive at the beta functions for the dimensionless couplings $\mathfrak{g}_i = k^4\bar{\mathfrak{g}}_i/Z_{\phi}^2$,
\begin{eqnarray}
	\beta_{\mathfrak{g}_1} &=& 2\,(2+\eta_\phi) \mathfrak{g}_1
	+ \frac{ \left(6 N_{\phi }+7\right) \mathfrak{g}_1^2 +  \left(N_{\phi }+15\right) \mathfrak{g}_2^2
		+2 \, \left(3 N_{\phi }+17\right) \mathfrak{g}_2 \mathfrak{g}_1 }{192 \pi ^2 N_{\phi }}  \nonumber \\
	&-& \left( \frac{10}{9 \pi  (1-2 \Lambda )^2} - 
	\frac{1}{9 \pi  (1-4 \Lambda/3)} - \frac{1}{18 \pi  (1-4 \Lambda/3)^2}\right)  
	\left(2 \mathfrak{g}_1+\mathfrak{g}_2\right) G \nonumber \\
	&-& \left( \frac{640 N_{\phi }}{27 (1-2 \Lambda )^3} - \frac{4 N_{\phi }}{27 (1-4 \Lambda/3)^2} - \frac{4 N_{\phi }}{27(1-4 \Lambda/3)^3} \right) G^2 \,, \label{eq:betas_O(N)_g1}
\end{eqnarray}
\begin{eqnarray}
	\beta_{\mathfrak{g}_2} &=& 2\,(2+\eta_\phi) \mathfrak{g}_2 + \frac{\left(N_{\phi }+21\right)\,\mathfrak{g}_2^2 +\mathfrak{g}_1^2+10 \mathfrak{g}_2 \mathfrak{g}_1}{192 \pi ^2 N_{\phi }} \nonumber \\
	&-&\left( \frac{5}{9 \pi  (1-2 \Lambda )^2} - \frac{1}{6 \pi  (3-4 \Lambda )} - \frac{1}{4 \pi  (3-4 \Lambda )^2} \right) (\mathfrak{g}_1+5 \mathfrak{g}_2) G \nonumber \\
	&+&\left( \frac{1280 N_{\phi }}{27 (1-2 \Lambda )^3} + \frac{4 N_{\phi }}{27 (1-4 \Lambda/3)^2} + \frac{4 N_{\phi }}{27(1-4 \Lambda/3)^3} \right) G^2 \,, \label{eq:betas_O(N)_g2}
\end{eqnarray}
with scalar anomalous dimensions given by
\begin{eqnarray}\label{eq:etas_O(N)}
	\eta_ \phi = \frac{(2N_\phi+1)\mathfrak{g}_1 + (N_\phi + 5)\mathfrak{g}_2 }{96\pi^2\, N_\phi} \,
	+ \, \left( \frac{1}{18 \pi (1-4 \Lambda/3)}+\frac{1}{18 \pi (1-4 \Lambda/3)^2}  \right) G\,.
\end{eqnarray}
In the absence of gravity, at $G = 0$, the systems features a Gaussian fixed point where $\mathfrak{g}_{1,*} = \mathfrak{g}_{2,*} = 0$. In the case without gravity, our analysis also shows indications for non-Gaussian fixed point solutions (cf.~App. \ref{App:Pure_Matter}). 

\begin{figure}[!t]
	\centering
	\hspace*{-.75cm}\includegraphics[height=7.0cm]{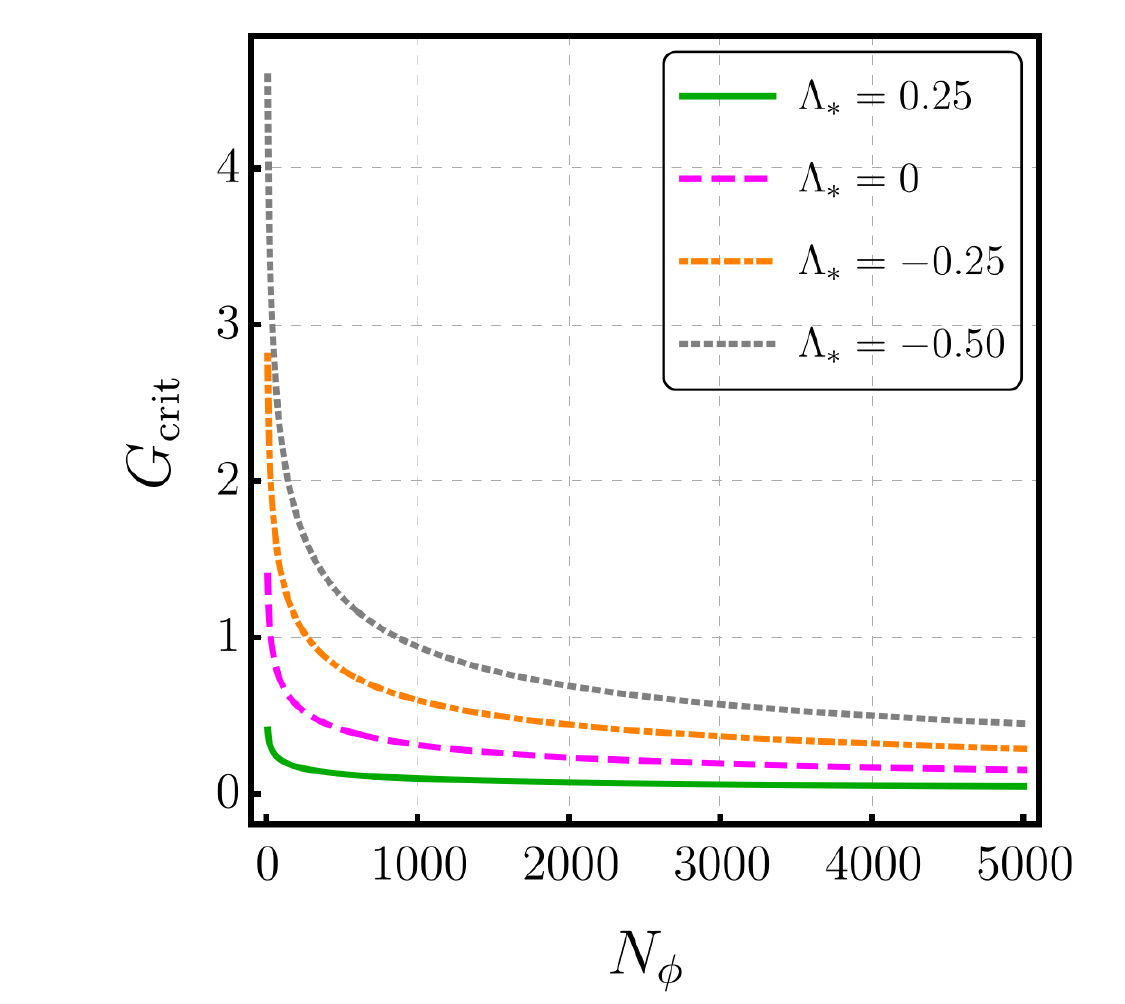}
	\caption{\label{fig:Gcrit_Nscalars} We show the weak-gravity bound as a function of $N_{\phi}$. $G_{\textmd{crit}}$ correspond to the value of $G_*$ at which the shifted Gaussian fixed point for $\mathfrak{g}_1$ and $\mathfrak{g}_2$ moves into the complex plane. The different curves correspond to different on the value of $\Lambda_*$.}
\end{figure}

In the presence of a non-Gaussian fixed point in the gravitational sector ($G_* \neq 0$), the system of flow equations \eqref{eq:betas_O(N)_g1}, \eqref{eq:betas_O(N)_g2} and \eqref{eq:etas_O(N)} does not support  a fixed point with vanishing values for $\mathfrak{g}_1$ and $\mathfrak{g}_2$. 

We expect that Eqs.~\eqref{eq:betas_O(N)_g1} and \eqref{eq:betas_O(N)_g2} could exhibit a weak-gravity bound, because both beta functions are parabolas with positive quadratic term which are shifted upwards by the gravitational term which is $\sim G^2$.
We confirm that the weak-gravity bound exists by analyzing both $\beta_{\mathfrak{g}_1}$ and $\beta_{\mathfrak{g}_2}$. We follow the shifted Gaussian fixed point as a function of $G_*$, $\Lambda_*$ and $N_\phi$ to check whether it becomes complex. For fixed values of $N_\phi$ and $\Lambda_*$, the weak-gravity bound is characterized by a critical value $G_{\textmd{crit}}$ at which the shifted Gaussian fixed point moves into the complex plane due to a fixed-point collision.

 In Fig.~\ref{fig:Gcrit_Nscalars}, we plot $G_{\textmd{crit}}$ as a function of the number of scalars for various choices of $\Lambda_*$. Our results indicate that the weak-gravity bound
\begin{itemize}
	\item  is present in a system with multiple scalar fields,
	\item  moves towards smaller $G_{\rm crit}$ when $N_{\phi}$ is increased,
	such that $G_{\textmd{crit}} \to 0$ in the large-$N_\phi$ limit,
	\item  follows a critical curve in the $G-\Lambda$ plane, such that $G$ increases with decreasing $\Lambda$.
\end{itemize}

\section{Upper bound on the number of scalars: Zero}\label{sec:upperbound}

In the previous sections, we have treated the gravitational couplings $G$ and $\Lambda$ as free parameters to delineate the weak-gravity boundary in the $G$-$\Lambda$ plane. We now determine the location of the fixed point with respect to this boundary. If the gravitational fixed-point values fall beyond the weak-gravity bound, then the joint gravity-scalar system does not exhibit asymptotic safety, at least within the present truncation. To perform this critical test of the asymptotic-safety paradigm, we combine our results from Sec.~\ref{sec:symmetries} with the fixed-point values obtained in the presence of matter fields.
For this discussion, it is relevant to distinguish universal and non-universal information: The fixed-point values are non-universal, as is the location of the weak-gravity bound. However, in combination they contain a universal piece of information: the statement that a fixed-point candidate lies below or above the
weak-gravity bound (i.e., the statement that the joint matter-gravity system has or does not have a fixed point) does not depend on the choice of scheme.

The gravitational fixed-point values can be calculated within distinct approximation schemes: the background-approximation preserves an auxiliary background diffeomorphism symmetry at the cost of making an approximation for the propagator of metric fluctuations. The fluctuation approach, reviewed in
\cite{Pawlowski:2020qer}, does away with this approximation, but does not presently account for symmetry identities that encode diffeomorphism symmetry. In the fluctuation approach, the Newton coupling is represented by several avatars which parameterize different interaction channels like the three-graviton and four-graviton coupling and are related by diffeomorphism symmetry.

Here, we report on results obtained in both settings. For the background approximation, we use the beta functions reported in \cite{Eichhorn:2017ylw}, which generalize \cite{Dona:2013qba} to our choice of gauge. For the fluctuation approach, we use the beta functions computed in Refs.~\cite{Eichhorn:2018ydy,Eichhorn:2018nda} and specialize to the 3-graviton avatar of the Newton coupling, assuming effective universality \cite{Eichhorn:2018akn,Eichhorn:2018ydy} to set all other avatars equal to it.

We neglect the presence of induced nonminimal matter-curvature couplings, because we assume  that they are subleading based on the analysis in \cite{Eichhorn:2017sok}. 

\begin{figure}[!t]
	\centering
	\hspace*{-1.cm}\includegraphics[height=7.0cm]{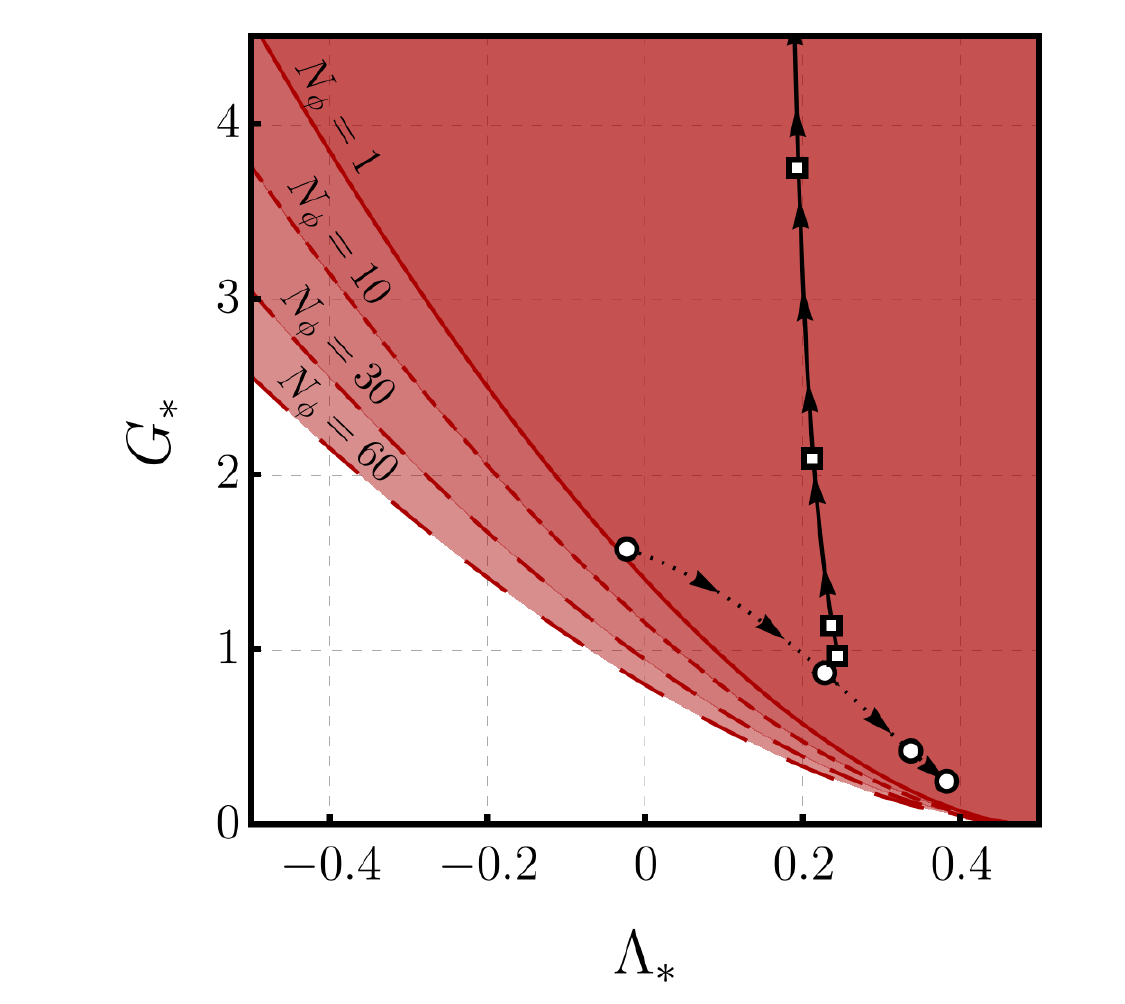}
	\caption{\label{fig:WGB-ManyScalars} In this plot we show the regions excluded by the weak-gravity bound for various choices of $N_\phi$. We also show the position of the gravitational fixed point, as a function of $N_\phi$, in both background- and fluctuation approach. The circle (square) markers correspond to fixed points evaluated using the background-approximation (fluctuation approach) with $N_\phi = 1$, $N_\phi=10$, $N_\phi = 30$ and $N_\phi = 60$ ($N_\phi = 1$, $N_\phi=10$, $N_\phi = 30$ and $N_\phi = 38$). The arrows indicate the direction of increasing $N_\phi$.}
\end{figure}

In Fig.~\ref{fig:WGB-ManyScalars}, we show the weak-gravity bound together with the fixed-point values  $G_{\ast}$ and $\Lambda_{\ast}$ for a varying number of scalars. For any number of scalars,  $G_{\ast}$ and $\Lambda_{\ast}$ lie in the excluded region for the two approximation schemes. The qualitative agreement of both approximation schemes hints at the robustness of our result. At the quantitative level, differences between both schemes persist, because the beta function for the cosmological constant differs among the two schemes \cite{Eichhorn:2018akn}.\\

The cosmological constant appears to play a key part in the above considerations, making it worthwhile to expand on its role:\\ 
First, the role and importance of the cosmological constant has been discussed from various points of view within asymptotic safety, e.g., \cite{Eichhorn:2013xr,Falls:2014zba,Percacci:2015wwa,Eichhorn:2017eht,Wetterich:2017ixo,Pagani:2020say}. Most recently the cosmological constant has been discussed as a candidate for an inessential coupling \cite{Baldazzi:2021orb}. As such, its fixed-point value would not appear in other beta functions directly, similar to the wave-function renormalization, which does not appear itself, instead, the associated anomalous dimension does. If such an understanding of the cosmological constant turned out to be correct, then the weak-gravity bound discussed here would be reduced to the weak-gravity bound in terms of essential couplings only, which for our present truncation would be only the Newton constant. Understanding structures like the weak-gravity bound in the light of discussions about the nature of the cosmological constant is an interesting direction for future work. \\
Second, we emphasize that the full graviton propagator contains further terms and couplings arising from higher-order curvature terms. Therefore, in an extended truncation, a combination of the cosmological constant with further couplings will appear in place of just the cosmological constant by itself. As an example, the weak-gravity bound in scalar-fermion systems has been investigated in a higher-order truncation in \cite{Eichhorn:2017eht}. Thus, even in the absence of the cosmological constant, the gravitational parameter space is expected to contain a nontrivial weak-gravity bound.

\begin{figure}[!t]
	\centering
	\hspace*{-.75cm}\includegraphics[height=6.5cm]{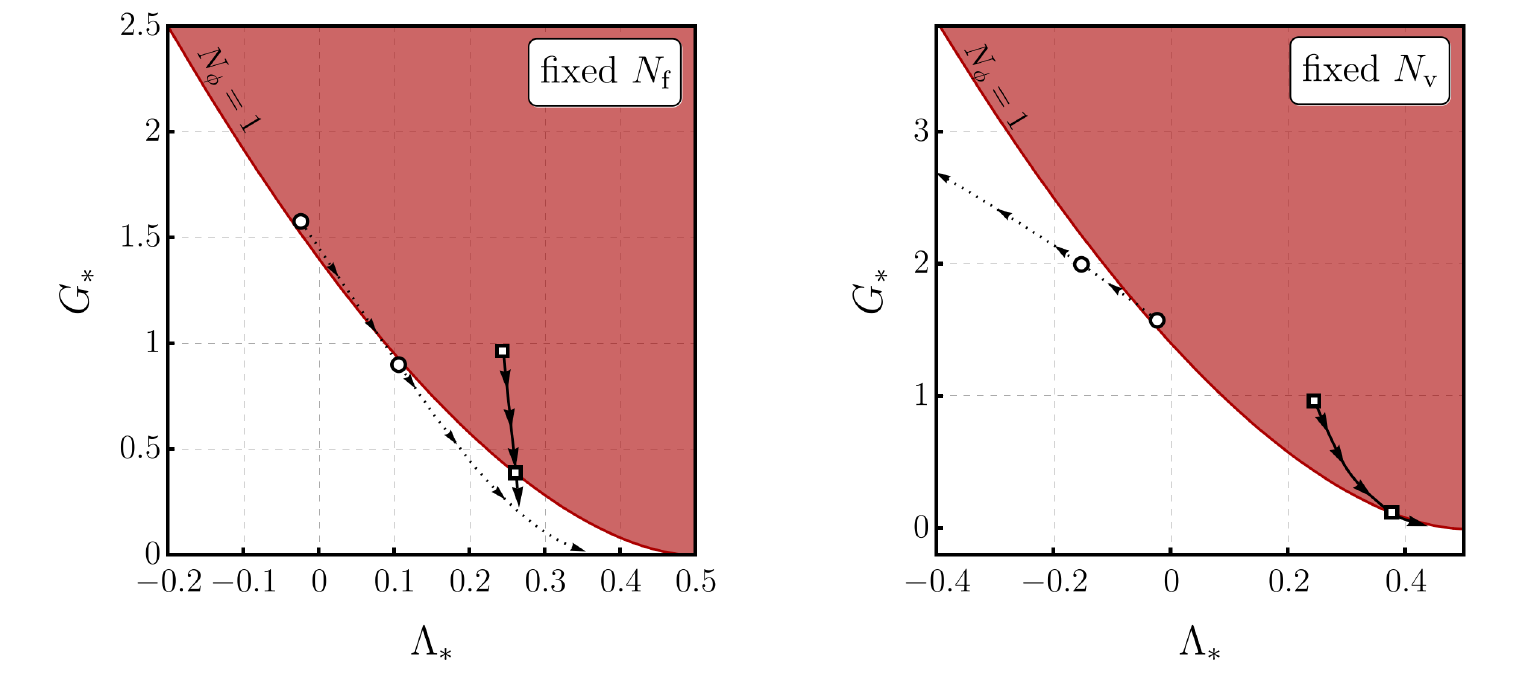}
	\caption{\label{fig:WGB+vec+ferm} Both plots show the region excluded by the weak-gravity bound in the case $N_\phi = 1$. In both cases, at a critical number of vector/fermion fields, the weak-gravity bound is evaded.
		Left panel: we show the position of the gravitational fixed point for different choices of $N_\textmd{v}$, while keeping the number of fermions fixed ($N_\textmd{f} = 0$). Right panel: we show the position of the gravitational fixed point for different choices of $N_\textmd{f}$, while keeping the number of vectors fixed ($N_\textmd{v} = 0$). The circle (square) markers correspond to fixed points evaluated using  the background-approximation (fluctuation approach) with $N_\textmd{v} = 0$ and $N_\textmd{v} = 3$ ($N_\textmd{v} = 0$ and $N_\textmd{v} = 242$) in the left panel and $N_\textmd{f} = 0$ and $N_\textmd{f} = 1/2$ ($N_\textmd{f} = 0$ and $N_\textmd{f} = 56$). The arrows indicate the direction of increasing $N_\textmd{v}$ or $N_\textmd{f}$.}
\end{figure}

The weak-gravity bound could be evaded, if additional matter fields of spin 1/2 or spin 1 are included. These additional fields drive the gravitational fixed-point values out of the excluded region, cf.~Fig.~\ref{fig:WGB+vec+ferm}. 
This effect could explain why the universe contains more than just gravity and scalar fields and vectors and spin 1/2 fields must also exist. In fact, the composition of the Standard Model might be required by a fundamental gravity-matter theory. We test this hypothesis by adding non-interacting vectors and spin 1/2 fermions to the scalar-gravity theory. 

For a theory with the same composition as the Standard Model, a real fixed point exists in the background-field approximation. \\
In contrast, additional vectors and/or fermions are required to achieve a real fixed point in the fluctuation-field approximation. This difference is because matter fields have a different effective ``weight’’ in the two approximations within the truncations that have been tested to date. In the fluctuation approximation, the ``weight” of a matter field, i.e., its contribution to the gravitational beta function, is highly suppressed compared to the ``weight” of metric fluctuations. This is not the case in the background approximation, where all degrees of freedom contribute with very similar ``weights”. \\
We conclude that in order to robustly establish the matter content required for a real fixed point to exist, truncations need to be extended. If it is found that matter fields beyond the Standard Model are required, these may constitute a dark sector. Such a dark sector, which would contain fermions and/or vectors, may have a distinct phenomenology. 

Our results complement studies which found that a gravitational fixed point exists under the impact of quantum fluctuations of Standard Model matter \cite{Dona:2013qba, Meibohm:2015twa,Biemans:2017zca,Alkofer:2018fxj,Wetterich:2019zdo}. For the first time, we find a mechanism that can simultaneously bound from below the number of fermion and vector fields that must exist. Additional constraints on the number of fermions exist if these are required to be light \cite{Gies:2018jnv,Gies:2021upb,deBrito:2020dta}. Taken together, these constraints indicate that requiring asymptotic safety may strongly constrain the structure and field content of a theory.

\section{Conclusions and outlook}\label{sec:conclusions}

In this paper, we have focused on the asymptotic-safety paradigm for scalar fields and gravity and explored it within a truncation of the gravity-matter dynamics. Our results are affected by systematic uncertainties due to the choice of truncation. With this caveat in mind, the following results hold:\\
In the asymptotic-safety paradigm, the gravitational coupling does not vanish. Therefore gravitational fluctuations couple to the scalar fields and generate scalar self-interactions. Our focus has been on these induced, \emph{unavoidable} scalar self-interactions. Their form is dictated by the global symmetries of the scalar field kinetic term, because the \emph{unavoidable} gravity-scalar vertices arise from the kinetic term and inherit its symmetries. These symmetries include a $\mathbb{Z}_2$ for each scalar ($\phi_i \rightarrow - \phi_i$), a shift-symmetry for each scalar ($\phi_i \rightarrow \phi_i +a_i$) and an O($N_{\phi}$) symmetry if $N_{\phi}$ real scalar fields are present. Therefore, in the presence of quantum gravity at finite coupling, it is not consistent to set these shift-symmetric interactions to zero. This result has been anticipated for $N_{\phi}=1$ in \cite{Eichhorn:2012va}  and here we have established it systematically at all $N_{\phi}$ within a truncation of the effective dynamics. 

Our findings have three implications for the asymptotic-safety paradigm.

First, we have found a weak-gravity bound in gravity-scalar systems. This bound (not a priori related to the similarly named weak-gravity conjecture) can arise whenever matter self-couplings are unavoidable under the impact of quantum gravity. In such a situation, the effective strength of quantum gravity has to be  weak in order to guarantee that a real-valued fixed point exists for the self-couplings. In the regime of strongly-coupled quantum gravity, no such fixed point remains.  Thus asymptotically safe gravity  must be sufficiently weakly coupled such that the joint scalar-gravity system may be asymptotically safe. The location of this weak-gravity bound in the parameter space of gravitational couplings is subject to systematic uncertainties that our choice of truncation introduces.

Second, the gravitational fixed-point values exceed the weak-gravity bound for any number of scalars. Thus, scalar-gravity systems cannot become asymptotically safe. This result should be understood within the systematic limitations of our study, namely our choice of truncation and Euclidean signature. We can therefore not fully exclude that there exist values for $N_{\phi}$ such that scalar-gravity systems become asymptotically safe. One might object that our result only holds under the assumption of maximum symmetry for the scalars, i.e., within a choice of universality class. To counter this objection, we highlight that scalar-gravity systems do not appear to feature additional universality classes: for instance, to the best of our knowledge there is no fixed point at finite scalar potential, i.e., with broken shift-symmetry. This situation can change, once additional matter degrees of freedom are present. The presence of gauge and Yukawa interaction, e.g., in the Standard Model, which can be nonzero in asymptotic safety, see, e.g., \cite{Harst:2011zx,Eichhorn:2016esv,Eichhorn:2017eht,Eichhorn:2017ylw,Eichhorn:2017lry}, may deform the weak-gravity bound. This deformation by non-zero gauge and Yukawa interactions is left for future work.

Third, the gravitational fixed-point values shift towards the weak-gravity regime, if additional matter fields (vectors and/or fermions) are present. 
We have found indications that the fixed-point value for the matter fields in the Standard Model could evade the weak-gravity bound. Accordingly, a universe with only gravity and scalar fields may not be consistent in the asymptotic-safety paradigm. 
Our result provides a step towards understanding the structure of the Standard Model in the asymptotic safety paradigm. The construction of a viable, asymptotically safe matter-gravity model appears to hinge on the presence of fermions and/or vectors. \\

Shift symmetry, which determines the interactions in our model, has been used extensively to construct models of dark energy. Many of those even feature a stronger symmetry, Galileon symmetry \cite{Nicolis:2008in,Curtright:2020cta}. It is presently an open question whether such dark-energy models have a natural place within the asymptotic-safety paradigm. This question warrants further investigation, given the unavoidable generation of shift-symmetric interactions. Building on functional Renormalization Group studies of Galileon systems without \cite{Steinwachs:2021jft} and with gravity \cite{Saltas:2016awg},  as well as non-renormalization theorems in Galileon systems without gravity \cite{Goon:2016ihr,Heisenberg:2019udf}, such an investigation is feasible.

 Shift symmetry by itself could also shape the properties of dynamical dark energy. Proposals for dark energy based on a scalar field with momentum-dependent interactions have been made, see, e.g., \cite{Langlois:2018dxi} for a review. If a scalar field is to be very light, this is naturally achievable in a setting with shift symmetry. Therefore, this is a potential testing ground for the hypothesis that global symmetries -- such as shift symmetry -- are realized in nature, see, e.g., \cite{Alvey:2021hjp}.\\

Additionally, our study complements an earlier study of the Higgs portal to scalar dark matter in asymptotic safety \cite{Eichhorn:2017als}. The standard Higgs portal to scalar dark matter vanishes under the impact of asymptotically safe gravity. Here, we have found that a dark scalar is coupled through a derivative portal. These interactions are parameterized by couplings of mass dimension -4, which remain irrelevant under the impact of asymptotically safe gravity. Accordingly, the values of these couplings are predictable at all scales, because UV scale symmetry fixes the values of irrelevant couplings at all scales. Due to their large negative mass dimension, the couplings run towards very small values in the infrared. It is to be expected that the corresponding interactions are too weak for the dark scalars to ever be in thermal equilibrium with the thermal bath of Standard Model fields in the early universe and thus thermal freezeout appears to be excluded. Alternative mechanisms for dark-matter production have been discussed, including the case of feebly-interacting dark matter, which can be produced out of thermal equilibrium. Whether the derivative couplings predicted by asymptotic safety could be large enough for such a production mechanism, remains an intriguing open question for the future.

Further, \cite{Garny:2015sjg} has proposed the idea of Planckian-interacting dark matter. The corresponding effective action features derivative interactions of the exact form we obtain here. Additionally, it relies on a mass term for the scalar. Mass terms for scalars are relevant in a large part of the asymptotically safe parameter space \cite{Narain:2009fy,Wetterich:2016uxm,Eichhorn:2017als,Pawlowski:2018ixd,Wetterich:2019rsn,Eichhorn:2020sbo}. Consequently, scalar masses few orders of magnitude below the Planck scale arise on RG trajectories without fine-tuning.
Accordingly, Planckian-interacting dark matter may find a natural explanation in the asymptotic-safety paradigm.

Finally, gravitational production on time-dependent backgrounds \cite{Ema:2018ucl,Chung:2018ayg}, which depends on the nonminimal coupling and the mass term, could be relevant for the present case. In this context, we stress that the nonminimal coupling is irrelevant in a large part of the gravitational coupling space \cite{Eichhorn:2020sbo} and therefore remains at zero at all scales. 

In summary, a dark matter sector consisting of a single scalar does not appear to be viable in asymptotic safety under the assumption of thermal production. Non-thermal mechanisms may be available, with Planckian-interacting dark matter being indicated as a promising idea due to its interaction structure which fits asymptotic safety naturally.\\

As a final point, we find it intriguing that arguments for the relative weakness of quantum gravity arise in distinct perspectives on gravity-matter systems, namely as the weak-gravity bound we discuss here and which has first been introduced in \cite{Eichhorn:2016esv,Christiansen:2017gtg, Eichhorn:2017eht} and as the weak-gravity conjecture which has first been introduced in \cite{ArkaniHamed:2006dz}. In \cite{deAlwis:2019aud}, conditions on an asymptotically safe fixed point were derived under which the weak-gravity conjecture would hold. It will be interesting to explore whether the weak-gravity bound and the weak-gravity conjecture can hold simultaneously in asymptotically safe gravity. In addition, understanding whether the relative weakness of quantum gravity might be a universal property that holds across distinct approaches to quantum gravity is an intriguing endeavor.

\acknowledgments
This work is supported by a research grant (29405) from VILLUM FONDEN. We thank M. Schiffer for discussions and M. Reichert for sharing data from \cite{Meibohm:2015twa}, which was used on early stages of this work. 

\appendix

\section{Functional RG methodology}\label{app:FRG}
In this section we present the Functional Renormalization Group (FRG) approach that we employ throughout this work. For reviews see, e.g., \cite{Pawlowski:2005xe,Gies:2006wv, Dupuis:2020fhh} and for reviews of the method in the context of quantum gravity, see, e.g., \cite{Eichhorn:2017egq,Percacci:2017fkn, Reuter:2019byg,Reichert:2020mja,Pawlowski:2020qer}. In this work we are using 
\begin{eqnarray}
	\int_x \equiv \int d^d x  \qquad \textmd{and} \qquad \int_p \equiv  \int \dfrac{d^d p}{(2\pi)^d},
\end{eqnarray}
to denote integration in $d$ space-like dimensions. In previous sections, we considered $d=4$.
\subsection{FRG and flow equation}
The FRG framework implements Wilson's idea \cite{Wilson:1973jj} of integrating out quantum fluctuations from high to low momentum modes in the path integral. To this end, we introduce an infrared cutoff scale $k$. Quantum fluctuations with momenta higher than $k$ are integrated out, whereas quantum fluctuations with momenta lower than $k$ are suppressed. This gives rise to an effective average (or flowing) action $\Gamma_k$ that interpolates between the UV and IR dynamics. $\Gamma_k$ agrees with the microscopic action at large $k$. While $k$ is lowered, the path integral is performed in a momentum-shell wise fashion, such that $\Gamma_{k\rightarrow 0}$ agrees with the full effective action. 
The main advantage of this setup is that instead of having to evaluate a functional integral, one can evaluate a functional differential equation. This flow equation encodes the change of $\Gamma_k$ as a function of $k$. The flow equation can be derived directly from the following definition of $\Gamma_k$,
\begin{eqnarray}
	\Gamma_k[\phi]=\int_x J \cdot \phi - \ln Z_k[J] - \Delta S_k[\phi] . \label{eq:defGamma}
\end{eqnarray}
The argument $\phi$ of $\Gamma$ is the expectation value of the quantum field $\varphi$, which constitutes the integration variable in the generating functional
\begin{eqnarray}
	Z_k[J] = \int_\Lambda \mathcal{D}\varphi \, e^{-S[\varphi]-\Delta S_k[\varphi]+\int J \cdot \varphi} .
\end{eqnarray}
Here we integrate only over field configurations $\varphi(p)$ with $|p^2|\leq\Lambda$ (UV regularization). The superfield $\varphi$ denotes the set of all fields to be taken into account (not only scalar fields). The IR regularization is implemented through the mass-like IR regulator term $\Delta S_k[\varphi]$  given by
\begin{eqnarray}
	\Delta S_k=\dfrac{1}{2}\int_p\varphi(-p)\textbf{R}_k(p)\varphi(p).
\end{eqnarray}
The regulator function $ R_k(p^2)$ must satisfy certain properties, namely,
\begin{eqnarray}
	&&\lim_{p^2/k^2\rightarrow 0} \textbf{R}_k(p) >0, \qquad \rm{(IR\:regularization)}, \\
	&&\lim_{k^2/p^2\rightarrow 0} \textbf{R}_k(p) = 0, \qquad \:\Gamma_{k\rightarrow 0}=\Gamma , \\
	&&\:\:\lim_{\substack{k^2 \rightarrow \Lambda\\ \textrm{fixed}\: p}}  \textbf{R}_k(p) \rightarrow \infty,  \:\:\:\:\:\:\:\:\: \Gamma_{k\rightarrow\Lambda} = S .
\end{eqnarray}
Apart from these requirements, the regulator function can be chosen freely, constituting different FRG schemes. This freedom affects non-universal quantities. Universal quantities are independent of this choice. From the definition \eqref{eq:defGamma}, we can derive the flow equation \cite{Wetterich:1992yh, Morris:1993qb}, pioneered for gravity in the seminal paper \cite{Reuter:1996cp}. For details of the derivation details see also the pedagogical accounts \cite{Gies:2006wv, Reichert:2020mja}. The flow equation is given by
\begin{eqnarray}
	\del_t \Gamma_k = \dfrac{1}{2}\Tr[(\Gamma_k^{(2)}+\textbf{R}_k)^{-1}\del_t \textbf{R}_k] , \label{flow}
\end{eqnarray}
where $\del_t = k \del_k $ and the superscript means two functional derivatives with respect to the components of the superfield. Accordingly, in a theory with more than one field, $\Gamma_k^{(2)}$ is a matrix in field space. Therefore the trace contains a summation in field space, as well as a summation over internal and spacetime indices as well as a summation/integration over the discrete/continuous eigenvalues of the (generalized) Laplacian that is part of $\Gamma_k^{(2)}$. The flow equation is a formally exact (non-perturbative) one-loop equation. Its one-loop structure does not mean that it only captures the first order in perturbation theory, because $\Gamma_k^{(2)}$ is not just the perturbative inverse propagator, see, e.g.,  \cite{Papenbrock:1994kf,Morris:1999ba} for early discussions of higher-loop effects and \cite{Dupuis:2020fhh} for numerous examples of non-perturbative physics. The formally exact nature of the flow equation is lost in practical applications, where systematic truncations of the space of couplings are necessary for practical calculations. The effective average action generically contains all quasilocal operators compatible with the choice of symmetries. The corresponding couplings make up the theory space, which is infinite-dimensional. For practical computational purposes, we truncate it. The other operators, which share the symmetries, are actually generated by the flow equation. As a consequence, the truncation is a source of systematic uncertainties and universal quantities may become dependent on the truncation scheme. Nonetheless, we work under the assumption that the back-coupling of these operators beyond the truncation is sub-leading. In the case of gravity(-matter) systems, a systematic truncation scheme follows the canonical mass dimension of couplings, because the scaling spectrum appears to be near-perturbative, see, e.g., \cite{Falls:2013bv,Falls:2014tra,Falls:2017lst,Falls:2018ylp,Eichhorn:2018ydy,Eichhorn:2018nda}.

For each running coupling $ \bar{g}_i =  \bar{g}_i  (k)$ of mass dimension $d_{\bar{g}_i}$ we define a dimensionless coupling, i.e., $g_i = \bar{g}_i\, k^{-d_{\bar{g}_i}}$. We focus on the beta functions of these dimensionless couplings,
\begin{eqnarray}
	\beta_{g_i}\equiv \del_t g_i.
\end{eqnarray}
For the wave function renormalization associated with the kinetic terms, $Z_i (\del_\mu \phi_i)^2$, we also define their anomalous dimensions by 
\begin{eqnarray}
	\eta_{\phi_i}\equiv -Z_{i}^{-1}\del_t Z_{i}.
\end{eqnarray}
RG trajectories start/end in \textit{fixed points} (FP), denoted by $g_*$ and defined by $
\del_t g_i (g_*) = 0 $, for all couplings in a given action. If all couplings satisfy $g_{i*}=0$, it is called a \textit{free (trivial or Gaussian) fixed point}. Otherwise, a fixed point is called a non-trivial, non-Gaussian or interacting fixed point. The Gaussian fixed point is guaranteed to exist in any field theory, whereas interacting fixed points are not guaranteed to exist.

In order to characterize the predictivity of a fixed point, we investigate the linearized flow in their vicinity. Intuitively, a fixed point can act as a sink (if RG trajectories emanate from it during the flow to the IR), as a source (if RG trajectories flow into it during the flow to the IR) or as a mix. Each direction in the space of couplings that is associated to sink is called irrelevant. It generates a prediction of the low-energy value of the coupling. Each direction that is associated to a source is called relevant. The low-energy value of such a (superposition of) couplings needs to be determined experimentally and constitutes a free parameter of the theory. The number of relevant directions is given by the number of positive critical exponents. The \textit{critical exponents} are the eigenvalues $\theta_i$ of the stability matrix, multiplied by an additional negative sign. The stability matrix is given by  
\be
M_{ij}=-\dfrac{\del \beta_{g_i} }{\del g_j}  \Big|_{g= g_{\ast}}. 
\ee 
If the real part of an eigenvalue is positive, it is associated with a \textit{relevant direction} (IR repulsive), whereas if it is negative, it is associated with an \textit{irrelevant direction} (IR attractive).

\subsection{PF expansion and choice of regulator function}

We employ a diagonal Litim regulator \cite{Litim:2002cf} that provides analytical flow equations and is given by a regulator function that depends on the wave-function renormalization of the regularized propagator. Additionally, it carries the same index structure as the inverse propagator, i.e.,
\begin{eqnarray}
	\textbf{R}_{k,\, i} = Z_{k,\,i}\, \Pi_i\, R_k(p^2),
\end{eqnarray}
where
\begin{eqnarray}
	R_k(p^2)=(k^2-p^2)\theta(k^2-p^2).\label{rk}
\end{eqnarray}
The label $i$ runs over the various fields and $\Pi_i$ denotes a spacetime tensor that carries the appropriate number of indices for the mode under consideration.

Before we can look for fixed points, we first need to find a consistent way to obtain the beta functions from the flow equation \eqref{flow}. This is achieved by projecting the field- and momentum-dependence of the rhs of Eq.~\eqref{flow} onto the different terms in our truncation. In a first step, we employ the $\mathcal{P}^{-1}\mathcal{F}$-expansion to isolate the relevant orders in a polynomial expansion in the fields. To that end, we define the following quantities:
\begin{eqnarray}
	\mathcal{P}_k&=&(\Gamma_k^{(2)}+\textbf{R}_k)\big|_ {\phi_i=0}, \label{eq:P} \\
	\mathcal{F}_k&=&\Gamma_k^{(2)}-\Gamma_k^{(2)}\big|_{\phi_i=0},\label{eq:F}
\end{eqnarray}
where $ \phi_i = 0 $ means that any remaining field dependence must be set zero. The first of these quantities is the free regularized propagator. $\mathcal{F}_k$ contains the interaction vertices. The flow equation can be expanded into a power series in $\mathcal{F}$. This is achieved by defining a derivative operator $\tilde\del_t$ that acts only on the $t$-dependence of the regulator $\textbf{R}_k$, through $\tilde\del_t=\int_p \del_t \textbf{R}_k \dfrac{\delta}{\delta \textbf{R}_k}$. Thus, we can rewrite the flow equation as
\begin{eqnarray}
	\del_t \Gamma_k \stackrel{\eqref{flow}}{=} \dfrac{1}{2}\Tr \tilde{\del}_t \ln \mathcal{P}_k+\dfrac{1}{2}\sum_{n=1}^{\infty}\dfrac{(-1)^{n+1}}{n}\Tr\tilde{\del}_t(\mathcal{P}^{-1}_k\mathcal{F}_k)^n , \label{pfflow}
\end{eqnarray}
such that the second term corresponds to an expansion in the number of vertices. The first term is a pure vacuum term that only contributes to the flow of the cosmological constant. To extract the anomalous dimension, we only need to take into account $n=1,2$ in Eq.~\eqref{pfflow} and to extract the beta functions for the quartic couplings, we need $n=1,2,3,4$. Whether or not contributions exist at these orders depends on the available vertices; e.g., if there are no three-point vertices, there is no contribution at order $n=4$.

\subsection{Shift and $\mathbb{Z}_2$ symmetric two-scalar-field model} \label{app:two_scalar}

We start with the following truncation with $k$-dependent couplings,
\begin{align}\label{eq:action2}
	\Gamma_k^{\textmd{2-Scalars}} =  \int_x 
	&\left( \frac{Z_{1}}{2} \pt_\mu \phi_1 \pt_\mu \phi_1 +
	\frac{Z_{2}}{2}  \pt_\mu \phi_2 \pt_\mu \phi_2  \right)  \\
	+ \int_x 
	&\bigg( \frac{\bar{g}_{1}}{8} \pt_\mu \phi_1 \pt_\mu \phi_1  \pt_\nu \phi_1 \pt_\nu \phi_1 + \frac{\bar{g}_{2}}{8}  \pt_\mu \phi_2 \pt_\mu \phi_2  \pt_\nu \phi_2 \pt_\nu \phi_2 \nonumber\\ 
	& + \frac{\bar{h}_{1}}{4}  \pt_\mu\phi_1\pt_\mu\phi_1 \pt_\nu \phi_2\pt_\nu \phi_2 + \frac{\bar{h}_{2}}{2}  \pt_\mu\phi_1\pt_\nu \phi_1 \pt_\mu \phi_2 \pt_\nu\phi_2  \bigg) , \nonumber
\end{align}
where the dimensionless counterparts of the renormalized couplings are given by
\begin{eqnarray}
	g_i =  k^{d} \bar{g}_i /Z_i^2 \qquad \textmd{and} \qquad h_i = k^{d} \bar{h}_i /(Z_1\, Z_2) ,
\end{eqnarray}
with $i=1,2$.
This action includes the leading-order shift-symmetric interactions of two scalars according to canonical power counting. The four couplings are canonically irrelevant. The first two ($g_1$ and $g_2$) are self-interaction quartic couplings, and the last two ($h_1$ and $h_2$) are quartic portal couplings, which mix $\phi_1$ and $\phi_2$. 
The action has two separate $\mathbb{Z}_2$ symmetries for each scalar field $\phi_{ 1,2}\rightarrow-\phi_{ 1,2}$, and two separate shift symmetries $\phi_{ 1,2} \rightarrow \phi _{ 1,2}+ a_{1,2 } $ (where $a_{1,2}$  are constants). The action is the most general action exhibiting these symmetries at quartic order in derivatives and fields. For special values of the couplings, an enhanced $\mathbb{Z}_4$ and O$(2)$ symmetry can be achieved. For any of these symmetries, a regulator function can be found that respects it. Consequently the flow equation will preserve the corresponding symmetry generating a flow that is confined to a subspace of the theory space. Under the coupling to gravity, the O$(2)$ symmetric subspace, which is two-dimensional, is the most relevant one, because no couplings outside this subspace are generated. We can work directly in this subspace of maximal symmetry \cite{Eichhorn:2017eht}, see Sec.~\ref{sec:symmetries}.

In the following, we work in Fourier space, because the couplings we are interested in are non-vanishing on a flat background. We compute $ \mathcal{P}_k$ taking two derivatives of the flowing action Eq.~\eqref{eq:action2}, setting both scalar fields in the resulting term to zero and adding the regulator. We obtain
\begin{eqnarray}
	\mathcal{P}_k^{-1}\left(p_1,p_2\right)_{\phi _1 \phi _1} =	\frac{(2\pi)^d\delta^d(p_1+p_2)}{Z_1 (p_1^2+R_k(p_1^2))}
	,\quad 
	\mathcal{P}_k^{-1}\left(p_1,p_2\right)_{\phi _2 \phi _2} = \frac{(2\pi)^d\delta^d(p_1+p_2)}{Z_2 (p_1^2+R_k(p_1^2))}.
	\label{eq:Ppurematter}
\end{eqnarray}
The off-diagonal elements $\mathcal{P}_{\phi_1\phi_2}$ and $\mathcal{P}_{\phi_2\phi_1}$ are zero.
Next, $\mathcal{F}_k$ can be computed according to its definition Eq.~\eqref{eq:F} by taking two derivatives of the flowing action \eqref{eq:action2} and then subtracting the expression for $\mathcal{P}$. 
Since the interaction terms in \eqref{eq:action2} are quartic, it follows that the elements of $ \mathcal{F}_k $ are quadratic in the fields. Explicitly, 
\begin{eqnarray}
	\mathcal{F}\left(p_1,p_2\right)_{ij}=\frac{1}{2}\sum_{a,b}\int _{q_1 q_2} f^{ij}_{ab}\left(q_{1,}q_{2;}p_{1,}p_2\right)\phi _a(q_1)\phi _b(q_2)(2 \pi )^d\delta^{d}( p_1+p_2+q_1+q_2),\quad
\end{eqnarray}
where $i,j,a,b=1,2$. The non-vanishing vertex elements are
\begin{align}
	f^{11}_{11,}\left(q_{1,}q_{2;}p_{1,}p_2\right)&=k^{-d}g_{1 } \left( Z_{1 }\right){}^2 q_{1 \mu } q_{2 \nu }  \left(\delta _{\mu \nu } p_{1 \alpha } p_{2\alpha }+p_{2\mu } p_{1 \nu }+p_{1 \mu } p_{2\nu }\right), \\
	f^{22}_{22}\left(q_{1,}q_{2;}p_{1,}p_2\right)&=k^{-d}g_{2 } \left( Z_{2 }\right){}^2q_{1 \mu } q_{2 \nu }  \left(\delta _{\mu \nu } p_{1 \alpha } p_{2\alpha }+p_{2\mu } p_{1 \nu }+p_{1 \mu } p_{2\nu }\right), \\
	f^{11}_{22}\left(q_{1,}q_{2;}p_{1,}p_2\right)&=k^{-d} Z_{1 } Z_{2 } q_{1 \mu } q_{2 \nu } \left(h_{1 } \delta _{\mu \nu } p_{1 \alpha } p_{2\alpha }+h_{2} \left(p_{2 \mu } p_{1\nu }+p_{1 \mu } p_{2\nu }\right)\right),\\
	f^{12}_{12}\left(q_{1,}q_{2;}p_{1,}p_2\right)&=k^{-d} Z_{1} Z_{2 }q_{1 \mu } q_{2 \nu } \left(h_{1 } p_{1 \mu } p_{2\nu }+h_{2}( p_{2\mu } p_{1 \nu }+\delta _{\mu \nu } p_{1 \alpha } p_{2\alpha })\right),\\
	f^{12}_{21}\left(q_{1,}q_{2;}p_{1,}p_2\right)&=f^{21}_{12}\left(q_{1,}q_{2;}p_{1,}p_2\right)=f^{21}_{21}\left(q_{2,}q_{1;}p_{1,}p_2\right)=f^{12}_{12}\left(q_{2,}q_{1;}p_{1,}p_2\right) ,\\
	f^{22}_{11}\left(q_{1,}q_{2;}p_{1,}p_2\right)&=f^{11}_{22}\left(q_{1,}q_{2;}p_{1,}p_2\right) .
\end{align}
The remaining terms are zero, namely
\begin{align}
	f^{12}_{11}\left(q_{1,}q_{2;}p_{1,}p_2\right)&=f^{12}_{22}\left(q_{1,}q_{2;}p_{1,}p_2\right)=f^{21}_{11}\left(q_{1,}q_{2;}p_{1,}p_2\right) = f^{21}_{22}\left(q_{1,}q_{2;}p_{1,}p_2\right)=0, \\
	f^{11}_{12}\left(q_{1,}q_{2;}p_{1,}p_2\right)&=f^{11}_{21}\left(q_{1,}q_{2;}p_{1,}p_2\right)=f^{22}_{12}\left(q_{1,}q_{2;}p_{1,}p_2\right) = f^{22}_{21}\left(q_{1,}q_{2;}p_{1,}p_2\right)=0.
\end{align}

\begin{figure}[!t]
	\hspace*{-2.8cm}\includegraphics[height=4cm]{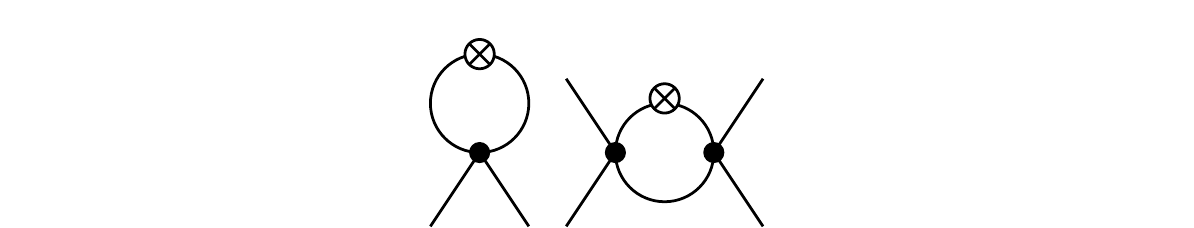}
	\caption{\label{fig:matter} We show the contribution to the anomalous dimension (left diagram) and the beta function of the scalar self-interaction (right diagram). Continuous lines represent the scalar fields 
		and the crossed circles represent regulator insertions.  
	}
\end{figure}

To extract the anomalous dimension $ \eta_i $, we divide both sides of \ref{pfflow} by $\phi_i(q_2)\phi_i(q_1)$ and set the fields to zero. The left hand side gives
\begin{eqnarray}
	\left(\frac{\delta ^2 \partial_t \Gamma _k}{ \delta \phi_i  (q_2) \delta \phi_i (q_1)}\right)_{\phi_1,\phi_2 =0}=-Z_{i }\eta _i  \,q_1^2(2 \pi )^d   \delta^{d}\left(q_1+q_2\right).  \label{lhseta}
\end{eqnarray}
For the right hand side we only need to consider the $n=1$ terms in the $\mathcal{PF}$-expansion since $ \mathcal{F}_k $ is quadratic in $\phi_i$: $\sum_{a}\int_{p_1,p_2}\left(\mathcal{P}^{-1}(p_2,p_1)_{\phi_a\phi_a}\mathcal{F}(p_1,p_2)_{\phi_a\phi_a}\right)$. This corresponds to tadpole diagrams, see Fig.~\ref{fig:matter}. It follows that
\begin{eqnarray}
	\hspace*{-.5cm}
	\left(\frac{\delta ^2 {\rm Tr} \,\tilde{\del}_t(\mathcal{P}^{-1}\mathcal{F})}{ \delta \phi_i  (q_2) \delta \phi_i (q_1)}\right)_{\phi_1,\phi_2 =0}
	\!\!= - \sum_{a}\int_p Z_{a}^{-2} f^{ii}_{aa}\left(-p,p,q_1,-q_1\right) G_{k}(p)^2 \del_t(Z_{a }R_k(p)) , \label{rhseta}
\end{eqnarray}
where $G_k(p)= (p^2+R_k(p))^{-1}$ and we have omitted an overall $ (2 \pi )^d \delta^{d}\left(q_1+q_2\right) $. After solving the last integral, comparing \eqref{lhseta} with \eqref{rhseta}, taking a derivative with respect to $q_1^2$, and taking into account the numerical factor in \eqref{pfflow}, we obtain 
\begin{eqnarray}
	\eta_i = \frac{ 1}{(4 \pi )^{\frac{d}{2}}(d+4) \Gamma  \left(\frac{d}{2}+1\right)}
	\left( g_{i } \left(d+4-\eta_i\right) + \frac{\left(d h_{1 }+2 h_{2 }\right) \left(d+4-\eta_j\right)}{d+2}\right), \label{etai}
\end{eqnarray}
where $j\neq i$. Next, to compute the beta functions for the couplings $g_{i}$ or $h_{i}$ we take four derivatives due to the quartic nature of the interaction terms. Thus we go up to second order in the vertex expansion in \eqref{pfflow}. As we have seen, the first order term is quadratic and does not contribute to the beta functions. For the second order, we only need to consider the contribution from candy diagram, see Fig.~\ref{fig:matter}. 

To obtain the beta functions for the quartic couplings, we take four derivatives with respect to the same field. The left hand side of the flow equation gives
\begin{align}\label{eq:LHSFlow_phi4}
	\left(\frac{\delta ^4 \del_t \Gamma_k}{\delta \phi _i (q_4)\delta \phi _i (q_3)\delta \phi _i (q_2) \delta \phi _i ( q_1)} \right)_{\phi_1,\phi_2=0}
	&= k^{-d} Z_{i}^2 \left(-d g_{i }+\beta _{g_{i }}-2 g_{i } \eta_i\right)   \\
	&\times(q_1 \cdot q_3 \, q_2 \cdot q_4 +q_1 \cdot q_4 \, q_3 \cdot q_2 + q_1 \cdot q_2 \,q_3 \cdot q_4) \nn \,.
\end{align}
The right hand side of the flow equation gives
\begin{align}\label{eq:RHSFlow_phi4}
	&\left(\frac{\delta ^4 \del_t \Gamma_k}{\delta \phi _i (q_4)\delta \phi _i (q_3)\delta \phi _i (q_2) \delta \phi _i ( q_1)} \right)_{\phi_1,\phi_2=0}  \\
	& =-\sum_{a}\int_{p_1,p_2}  \!\!Z_{a }^{-2} f^{ii}_{aa}\left(-p_1,p_2,q_1,-q_1\right) f^{ii}_{aa}\left(-p_2,p_1,q_3,-q_3\right) G_k(p_1)  G_k(p_2)^2  \del_t(Z_{a } R_k(p_2)) ,\nn
\end{align}
which comes from the $(\mathcal{P}^{-1}  \mathcal{F})^2$-contribution. Here we already neglect terms at higher order in momenta, which are generated, but not part of the interactions we include in our truncation.
Both in Eqs.~\eqref{eq:LHSFlow_phi4} and \eqref{eq:RHSFlow_phi4} we omitted the factor $ (2 \pi )^d \delta^{ d }  \left(q_1+q_2+q_3+q_4\right)$. By comparing both sides of the flow equation, we obtain
\begin{eqnarray}
	\beta _{g_1}&=&g_1 \left(d+2 \eta_1\right)+\frac{16 }{(4 \pi )^{\frac{d}{2}}\Gamma \left(\frac{d}{2}\right)}  \\
	& \times&\left( \frac{\left(d^2+6 d+20\right) \left(d-\eta_1+6\right)\,g_1^2}{4d (d+2) (d+4) (d+6)} +  \frac{\left(d-\eta_2+6\right)}{(d+4) (d+6)} \left(\frac{h_1 h_2}{d}+\frac{3 h_2^2}{d (d+2)}+\frac{h_1^2}{4}\right) \right), \nn\\
	\beta_{g_2}&=&g_2 \left(d+2 \eta_2\right)+ \frac{16 }{(4 \pi )^{\frac{d}{2}}\Gamma \left(\frac{d}{2}\right)} \\
	& \times&\left( \frac{\left(d^2+6 d+20\right) \left(d-\eta_2+6\right)\,g_2^2}{4d (d+2) (d+4) (d+6)} +  \frac{\left(d-\eta_1+6\right)}{(d+4) (d+6)} \left(\frac{h_1 h_2}{d}+\frac{3 h_2^2}{d (d+2)}+\frac{h_1^2}{4}\right) \right). \nn
\end{eqnarray}

For the portal beta functions, we take two derivatives with respect to $\phi_1$ and then two with respect to $\phi_2$. To disentangle the two different couplings at this order in the fields and second order in derivatives, we need to specify a momentum configuration (that respects momentum conservation). We start with
\begin{eqnarray}
	&{}&\left( \frac{\delta ^4 \del_t \Gamma_k}{\delta \phi _2 (q_4)\delta \phi _2 (q_3)\delta \phi _1 (q_2) \delta \phi _1 (   q_1)} \right)_{\phi_1,\phi_2=0} \\
	&{}& \qquad\qquad\qquad\qquad =k^{-d} Z_{1} Z_{2 } \,q_1\cdot q_2 \, q_3\cdot q_4 \, \left( \beta _{h_1} -h_1 \left( d+\eta_1+\eta_2 \right) \right)  \nn \\
	&{}& \qquad\qquad\qquad\qquad \,+\, k^{-d} Z_{1} Z_{2 } \left( q_1\cdot q_3 \, q_2\cdot q_4+q_1\cdot q_4 \,q_3\cdot q_2 \right)  \left(\beta _{h_2}-h_2 \left(d+\eta_1+\eta_2\right)\right) . \nn
	\label{eq:ddtGammaportal}
\end{eqnarray}
To project onto $\beta_{h_1}$, we choose $q_1 = - q_2$, $q_3 = - q_4$ and $q_1 \cdot q_3 = 0$. To project onto $\beta_{h_2}$, we choose $q_1 = - q_3$, $q_2 = - q_4$ and $q_1 \cdot q_2 = 0$.
We thereby obtain
\begin{eqnarray}
	\hspace*{.5cm}\,\,\,\,
	\beta _{h_1}&=& \left(d+\eta _1 + \eta _2 \right) h_{1} + \frac{2}{(4 \pi )^{\frac{d}{2}}\Gamma \left(\frac{d}{2}\right)}  \\
	&\times& \Bigg[  \left(\frac{h_{1}+h_{2}}{d}+\frac{2 h_{2}}{d (d+2)}+\dfrac{h_{1} }{2}\right) \left( \frac{2(d+6-\eta_1) g_1 }{(d+4) (d+6)} + \frac{2(d+6-\eta_2) g_2 }{(d+4) (d+6)} \right)  \nn\\
	&{}& + \frac{(h_1 + h_2)^2}{d(d+2)}  \frac{8 d + 12 - 4 \eta_1 - 4\eta_2}{(d+4) (d+6)} \Bigg] \,,\nn
\end{eqnarray}
\begin{eqnarray}
	\hspace*{-1.7cm}
	\beta _{h_2}&=& \left(d+\eta _1 + \eta _2 \right)h_{2} + \frac{2}{(4 \pi )^{\frac{d}{2}}\Gamma \left(\frac{d}{2}\right)}  \\
	&\times& \Bigg[  \frac{2h_2}{d(d+2)} \left( \frac{2(d+6-\eta_1) g_1 }{(d+4) (d+6)} + \frac{2(d+6-\eta_2) g_2 }{(d+4) (d+6)} \right)  \nn\\
	&{}& + \left(\frac{\left(h_{1} +h_{2}\right){}^2}{d (d+2)}+\frac{h_{2} \left(h_{1} +h_{2}\right)}{d}+ \frac{h_{2}^2}{2}\right)   \frac{8 d + 12 - 4 \eta_1 - 4\eta_2}{(d+4) (d+6)} \Bigg] \,.\nn 
\end{eqnarray}

There are several common approximations regarding the inclusion of the anomalous dimensions in Eq.~\eqref{etai}. The first one is to replace the full expression for $\eta_i$ in the expressions for the beta functions:
This can be interpreted as a resummation of parts of a perturbative expansion, as coupling appear in the denominator of the beta functions. On the other hand, these same terms can introduce artificial poles and zeros in the beta functions. The second option is to set $\eta_i=0$, where it arises from the regulator insertion $\partial_t \textbf{R}_k$. This perturbative approximation reproduces universal one-loop results. It amounts to taking the leading order terms in a Taylor expansion of Eq.~\eqref{etai} in the couplings into account in the ``canonical" terms in the beta function, but not in the loop contributions.
We use this approximation throughout this work, as it enables an analytical search for fixed-point solutions.

\subsection{Coupling to gravity}
\begin{figure}[!t]
	\hspace*{-2.8cm}\includegraphics[height=4cm]{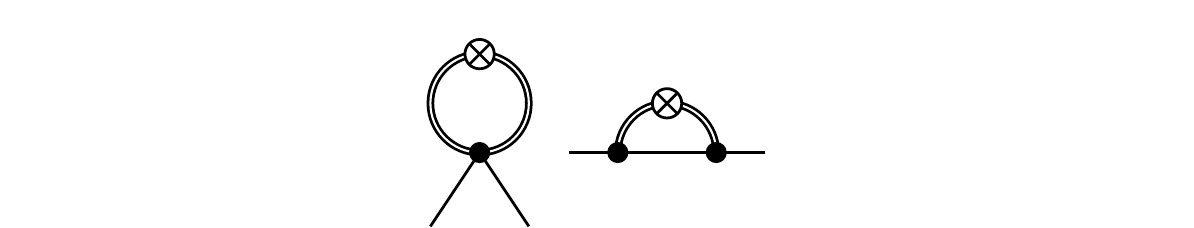}
	\caption{\label{fig:diagramsgravity} We show the gravitational contribution to the anomalous dimension of the scalar. Double lines represent metric fluctuations, single lines scalars. The regulator insertion, denoted by a crossed circle, must appear on each of the internal lines in turn, thus there are two versions of the sunset-diagram.}
\end{figure}

\begin{figure}[!t]
	\centering
	\hspace*{-1.6cm}\includegraphics[height=6cm]{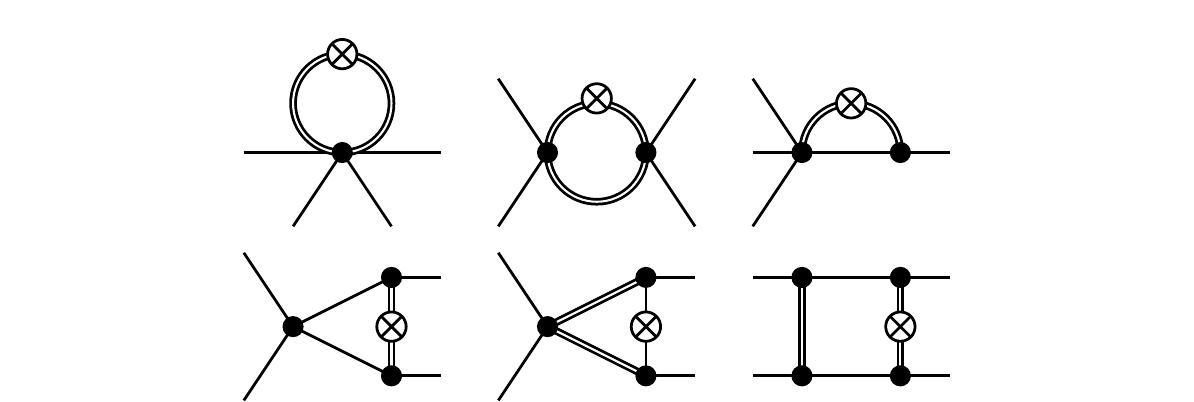}
	\caption{\label{fig:diagramsgravity2} These diagrams contribute to the beta functions of the quartic couplings with gravity. }
\end{figure}

In this section we extend our previous model to the presence of a fluctuating metric. To set up a coarse-graining flow, we need a notion of locality that allows us to distinguish UV modes from IR modes. We implement this notion of locality by choosing a background metric $\bar{g}_{\mu\nu}$ and making use of the background-field method \cite{Reuter:1996cp}. For a discussion of how background independence can be achieved in the presence of such a background, see \cite{Becker:2014qya,Morris:2016spn,Percacci:2016arh,Ohta:2017dsq,Eichhorn:2018akn,Pawlowski:2020qer}. For our purposes, we can work with the technically simplest choice of background metric, namely $\bar{g}_{\mu\nu} = \delta_{\mu\nu}$, a flat Euclidean background. This allows us to extract the beta functions in the matter sector unambiguously. Here, the choice of signature is fixed by the FRG framework (and is one of the technical limitations of this work), but the choice of vanishing background curvature is the easiest choice to disentangle the couplings we are interested in from similar non-minimal matter-gravity interactions. Accordingly, we write the spacetime metric in terms of a fluctuating metric $h_{\mu\nu}$ around an Euclidean background as
\begin{eqnarray}
	g_{\mu\nu}=\delta_{\mu\nu}+\kappa h_{\mu\nu}, \qquad \kappa = \sqrt{32\pi G_{N} Z_h}. \label{background}
\end{eqnarray}
Here, $G_{N}$ and $ Z_h$ are the Newton coupling and the wave function renormalization for the fluctuation field $h_{\mu\nu}$. We now choose our truncation as 
\begin{eqnarray}
	\Gamma_k [g_{\mu\nu},\phi_i]= \Gamma^\textmd{matter}_k+ \Gamma^{\rm EH}_k,
\end{eqnarray}
where $\Gamma^\textmd{matter}_k$ is the matter sector, see Sec.~\ref{App:Pure_Matter}, coupled with gravity through \eqref{background}. $\Gamma^{\rm EH}_k$ is the Einstein-Hilbert action where
$G= G_N\, k^{d-2}$ and $\Lambda=\bar{\Lambda} k^{-2} $ are the dimensionless couplings associated with the Newton coupling and the cosmological constant:
\begin{eqnarray}
	\Gamma^{\rm EH}_k=-\dfrac{1}{16\pi G_{N}}\int_x \sqrt{g}(R-2\bar{\Lambda}).
\end{eqnarray}
All vertices relevant for our work arise from an expansion up to second order in $h_{\mu\nu}$. 
We need to include a gauge fixing term so that we can obtain the propagator for the fluctuation field:
\begin{eqnarray}
	S_{\rm{gf}} = \dfrac{1}{\alpha}\int_x  \delta^{\mu\nu}F_{\mu}F_\nu, \qquad F_\mu=\del^\nu h_{\mu\nu}-\dfrac{\beta+1}{d}\del_\mu h,
\end{eqnarray}
where $\alpha$ and $\beta$ are gauge-fixing parameters that we choose as $\alpha \to 0$ and $\beta \to 0$.

We decompose the rank-2 symmetric tensor $h_{\mu\nu}$ into its irreducible components under the $d$-dimensional rotation group through the York decomposition
\begin{eqnarray}\label{eq:YorkDecomp}
	h_{\mu\nu}=\htt_{\mu\nu}+2\del_{(\mu}V_{\nu)}+\left(\del_\mu\del_\nu-\dfrac{1}{d}\delta_{\mu\nu}\nabla^2 \right) \sigma + \dfrac{1}{d}\delta_{\mu\nu} h. 
\end{eqnarray}
Here $\htt$ is a transverse traceless tensor that corresponds to the helicity two mode, $V$ is a transverse vector (helicity one), $\sigma$ and $h$ are scalar fields (helicity zero). The last one is just the trace of $h_{\mu\nu}$. In our choice of gauge the transverse vector and the scalar $\sigma$ contribution to the beta functions of matter couplings vanishes. The effect of different choices of $\beta$ is discussed in App. \ref{app:GaugeDep}.

The Litim regulator for the TT and the trace mode is
\begin{eqnarray}
	\textbf{R}^{\textmd{TT}}_k(p^2)_{\mu\nu\alpha\beta}&=&Z_h (k^2-p^2)\theta(k^2-p^2)P^{\textmd{TT}}_{\mu\nu\alpha\beta}(p^2) ,\\
	\textbf{R}^{\textmd{tr}}_k(p^2)&=& -\dfrac{(d-2)(d-1)}{d^2}Z_h(k^2-p^2)\theta(k^2-p^2).
\end{eqnarray}
For the TT mode, $ P^{\textmd{TT}}_{\mu\nu\alpha\beta}=\dfrac{1}{2}(\Theta_{\mu\alpha}\Theta_{\nu\beta}+\Theta_{\mu\beta}\Theta_{\nu\alpha})-\dfrac{1}{d-1}\Theta_{\mu\nu}\Theta_{\alpha\beta} $ is the spin-2 projection operator, where $\Theta_{\mu\nu}(p)=\delta_{\mu\nu}-\dfrac{p_\mu p_\nu}{p^2}$.

The presence of metric fluctuations results in additional contributions to the beta functions in the matter sector, shown in Fig.~\ref{fig:diagramsgravity} and Fig.~\ref{fig:diagramsgravity2}. The explicit expressions for the anomalous dimensions and beta functions can be found in an auxiliary file. 

To evaluate the flow of RG-equations presented in this work we use the self-written Mathematica codes based on the packages \textit{xAct}  \cite{Brizuela:2008ra,Martin-Garcia:2007bqa,MartinGarcia:2008qz}, \textit{DoFun} \cite{Huber:2011qr,Huber:2019dkb}, \textit{FormTracer} \cite{Cyrol:2016zqb} and \textit{Package-X} \cite{Patel:2016fam}.

\section{Shift-symmetric fixed-point candidates without gravity}\label{App:Pure_Matter}

In this appendix, we summarize our main results for the fixed-point structure of shift-symmetric scalar-field models in a pure-matter setting, i.e., without gravity.

\subsection{Two-scalar fields model}

We start from a shift-symmetric model involving two real scalar fields, with the truncation defined by \eqref{Truncation_2scalar}. In addition to shift-symmetry, the flowing action \eqref{Truncation_2scalar} is also invariant under $\mathbb{Z}_2 \times \mathbb{Z}_2$ transformations. In the pure-matter setting, the truncated flow is described by the beta functions
\begin{eqnarray}
	\beta_{g_1}& =& (4+2\eta_1)\,g_1 + \frac{5g_1^2 + 2h_1^2 + h_2^2+2h_1h_2}{64\pi^2}  \,,\label{eq:betag1_pure_matter}\\
	\beta_{g_2}& =& (4+2\eta_2)\,g_2 + \frac{5g_2^2 + 2h_1^2 + h_2^2+2h_1h_2}{64\pi^2}  \,,\label{eq:betag2_pure_matter}\\
	\beta_{h_1} &=& (4+\eta_1+\eta_2)\, h_1 + \frac{(g_1+g_2)(9h_1+4h_2)+(h_1+h_2)^2}{192\pi^2}  \,,\label{eq:betah1_pure_matter} \nonumber\\ &{}&\\
	\beta_{h_2} &=& (4+\eta_1+\eta_2)\, h_2 + \frac{h_1^2+8h_1h_2 + 19 h_2^2 + h_2(g_1+g_2)}{192\pi^2}  \,,\nonumber\\ &{}&\label{eq:betah2_pure_matter}
\end{eqnarray}
with anomalous dimensions
\begin{eqnarray}
	\eta_1 = \frac{3g_1 + 2h_1+h_2}{96\pi^2} 
	\qquad \textmd{and} \qquad 
	\eta_2 = \frac{3g_2 + 2h_1+h_2}{96\pi^2} \,.\label{eq:anomdim2_pure_matter}
\end{eqnarray}
Besides the Gaussian fixed point ($g_{1 *} = g_{2 *} = h_{1 *} = h_{2 *} = 0$), within our truncation we have found additional interacting fixed-point candidates, cf.~Table \ref{tab:FPs_2scalars}. We refer to zeros of the beta functions as fixed-point candidates, when the deviation from canonical scaling is so large that canonically very irrelevant couplings (of mass dimension $-4$) are shifted into relevance. Such large shifts contradict our truncation scheme, which assumes near-canonical scaling. Fixed points with different scaling can nevertheless be found with our truncation, but not described robustly. Therefore, we carefully refer to these as fixed-point candidates.

\begin{table}[t]
	\begin{center}
		\begin{tabular}{|c|c|c|c|c|c|c|c|c|c|c|}
			\hline\hline 
			& $g_{1 *}$ & $g_{2 *}$ & $h_{1 *}$ & $h_{2 *}$ & $\eta_{1 *}$ & $\eta_{2 *}$ & $\theta_1$ & $\theta_2$ & $\theta_3$ & $\theta_4$\\ \hline\hline
			$\!\textmd{\textbf{FP}}_{1,1}\!$ & -64.17 & -159.72 & -95.31 & -193.07 & -0.61 & -0.91 & 4.00 & 0.91 & -0.70 & -2.60 \\
			$\!\textmd{\textbf{FP}}_{1,2}\!$ & -159.72 & -64.17 & -95.31 & -193.07 & -0.91 & -0.61 & 4.00 & 0.91 & -0.70 & -2.60 \\
			$\!\textmd{\textbf{FP}}_{1,3}\!$ & 0.00 & -280.74 & 0.00 & 0.00 & 0.00 & -0.89 & 4.00 & -1.78 & -2.96 & -4.00 \\
			$\!\textmd{\textbf{FP}}_{1,4}\!$ & -280.74 & 0.00 & 0.00 & 0.00 & -0.89 & 0.00 & 4.00 & -1.78 & -2.96 & -4.00 \\\hline\hline
			$\!\textmd{\textbf{FP}}_{2,1}\!$& -140.37 & -140.37 & -140.37 & -140.37 & -0.89 & -0.89 & 4.00 & \underline{0.89} & 0.44 & -1.93 \\
			$\!\textmd{\textbf{FP}}_{2,2}\!$& -280.74 & -280.74 & 0.00 & 0.00 & -0.89 & -0.89 & 4.00 & \underline{4.00} & 0.44 & -1.93 \\
			$\!\textmd{\textbf{FP}}_{2,3}\!$& -311.18 & -311.18 & 190.38 & -5.91 & -0.59 & -0.59 & \underline{4.08} & 4.00 & -0.84 & -3.12 \\
			$\!\textmd{\textbf{FP}}_{2,4}\!$& -66.31 & -66.31 & -54.49 & -250.78 & -0.59 & -0.59 & 4.00 & -0.84 & \underline{-1.35} & -3.12 \\\hline\hline
			$\!\textmd{\textbf{FP}}_{3,1}\!$& -195.30 & -195.30 & -165.69 & -14.81 & -0.98 & -0.98 & 4.00 & \underline{2.30}$^*\!$ & -0.61$^*\!$ & -1.18 \\ \hline\hline
		\end{tabular}
		\caption{Fixed-point candidates obtained in a shift-symmetric model with two scalar fields. The fixed points are separated into three different categories, defined with respect to enhanced symmetry. In addition to the fixed point values, we also report the corresponding anomalous dimensions and critical exponents.}
		\label{tab:FPs_2scalars}
	\end{center}
\end{table}

The set of fixed-point candidates can be separated into three different categories according to their symmetries enhancement properties.

The first category, containing $\textmd{\textbf{FP}}_{1,1}$, $\textmd{\textbf{FP}}_{1,2}$, $\textmd{\textbf{FP}}_{1,3}$ and $\textmd{\textbf{FP}}_{1,4}$, is defined by the lack of symmetry enhancement. The fixed-point candidates in this category appear in pairs that can be identified by exchanging $g_1 \leftrightarrow g_2$, i.e.
\begin{eqnarray}
	\textmd{\textbf{FP}}_{1,1} = \textmd{\textbf{FP}}_{1,2}|_{g_1 \leftrightarrow g_2}  \qquad \textmd{and} \qquad \textmd{\textbf{FP}}_{1,3} = \textmd{\textbf{FP}}_{1,4}|_{g_1 \leftrightarrow g_2}  \,.
\end{eqnarray}
This property is expected from the RG flow, cf.~\eqref{eq:betag1_pure_matter}, \eqref{eq:betag2_pure_matter}, \eqref{eq:betah1_pure_matter}, \eqref{eq:betah2_pure_matter} and \eqref{eq:anomdim2_pure_matter}, since the system of beta functions remains invariant under the exchange $g_1 \leftrightarrow g_2$.

The  fixed-point candidates $\textmd{\textbf{FP}}_{1,1}$ and $\textmd{\textbf{FP}}_{1,2}$ are fully non-Gaussian, i.e., all the couplings in our truncation are non-vanishing. They have two IR repulsive directions. The corresponding critical exponents deviate by a large amount from the canonical dimension. Accordingly, our truncation allows at best a qualitative study of these fixed points, but more extended truncations would be required to establish more robustly that these fixed points exist.

Meanwhile, the fixed-point candidates $\textmd{\textbf{FP}}_{1,3}$ and $\textmd{\textbf{FP}}_{1,4}$ are partially Gaussian, i.e., some of the couplings vanish at the fixed point. They are trivial extensions of fixed-point candidates in a shift-symmetric truncation with a single scalar field. Therefore they inherit the single IR-repulsive direction from the fixed-point candidates with a single field.

The second category, containing $\textmd{\textbf{FP}}_{2,1}$, $\textmd{\textbf{FP}}_{2,2}$, $\textmd{\textbf{FP}}_{2,3}$ and $\textmd{\textbf{FP}}_{2,4}$, is defined by the symmetry enhancement $\mathbb{Z}_2 \times \mathbb{Z}_2 \rightarrow \mathbb{Z}_4$. Fixed-point candidates belonging to this category are characterized by the relations 
\begin{eqnarray}\label{Z4_relations}
	g_1 = g_2 \qquad \textmd{and}\qquad \eta_1 = \eta_2 \,.
\end{eqnarray}
Thus the fixed point action is invariant under 
\begin{eqnarray}
	\begin{pmatrix}
		\phi_1 \\
		\phi_2
	\end{pmatrix}
	\mapsto
	\begin{pmatrix}
		\cos(n \pi /2) & -\sin(n  \pi /2) \\
		\sin(n \pi/2) & \cos(n \pi/2)
	\end{pmatrix}
	\begin{pmatrix}
		\phi_1 \\
		\phi_2
	\end{pmatrix} \,,\quad \textmd{with}\,\, n \in \mathbb{Z}\,.
\end{eqnarray}
Within this category, only $\textmd{\textbf{FP}}_{2,2}$ exhibits a partially Gaussian nature. At this fixed-point candidate the portal couplings $h_1$ and $h_2$ are zero. This fixed-point candidate is a trivial generalization of fixed-point candidates in single-scalar models. Both $\textmd{\textbf{FP}}_{2,1}$ and $\textmd{\textbf{FP}}_{2,2}$ features three IR repulsive directions. The fixed point candidates $\textmd{\textbf{FP}}_{2,3}$ and $\textmd{\textbf{FP}}_{2,4}$ exhibit two and one IR repulsive directions, respectively. The critical exponents that are underlined in Tab.~\ref{tab:FPs_2scalars} correspond to eigenvectors lying outside of the $\mathbb{Z}_4$-subspace. For the fixed points $\textmd{\textbf{FP}}_{2,1}$, $\textmd{\textbf{FP}}_{2,2}$ and $\textmd{\textbf{FP}}_{2,3}$ the marked critical exponents are positive, which implies that the correspondent directions are IR repulsive. Accordingly, generic initial conditions do not result in a symmetry enhancement to $\mathbb{Z}_4$- symmetry under the RG flow to the IR. For the fixed point $\textmd{\textbf{FP}}_{2,4}$ the marked critical exponent is negative, indicating that the corresponding direction is IR attractive such that this fixed point can be reached by trajectories starting outside the $\mathbb{Z}_4$-subspace.

The third category, containing only the fixed point $\textmd{\textbf{FP}}_{3,1}$, is defined by the additional symmetry enhancement $\mathbb{Z}_4 \rightarrow \textmd{O}(2)$. Fixed-point candidates in this category are characterized by
\begin{eqnarray}\label{O(2)_relations}
	g_1 = g_2\,,\qquad g_1 - h_1 - 2h_2 = 0 \qquad \textmd{and} \qquad \eta_1 = \eta_2\,.
\end{eqnarray}
At the level of the fixed point action, the enhanced symmetry corresponds to field transformations of the form
\begin{eqnarray}
	\begin{pmatrix}
		\phi_1 \\
		\phi_2
	\end{pmatrix}
	\mapsto
	\begin{pmatrix}
		\cos(\theta) & -\sin(\theta) \\
		\sin(\theta) & \cos(\theta)
	\end{pmatrix}
	\begin{pmatrix}
		\phi_1 \\
		\phi_2
	\end{pmatrix} \,,\quad \textmd{with}\,\, \theta \in \mathbb{R}\,.
\end{eqnarray}
We find two IR-repulsive directions associated with the fixed point $\textmd{\textbf{FP}}_{3,1}$. The negative critical exponent marked by an asterisk corresponds to an eigenvector lying outside the O$(2)$-subspace, but still contained in the  $\mathbb{Z}_4$-subspace. The positive critical exponent that is underlined and marked by an asterisk is associated with an eigenvector that breaks both O$(2)$- and $\mathbb{Z}_4$-symmetries.

The fixed point candidates lie in symmetry-enhanced subspaces of the shift-symmetric theory space. Eqs.~\eqref{Z4_relations} and \eqref{O(2)_relations} are not only valid at the fixed point, but also preserved along the flow.
This property can be verified explicitly in the flow equations \eqref{eq:betag1_pure_matter}, \eqref{eq:betag2_pure_matter}, \eqref{eq:betah1_pure_matter}, \eqref{eq:betah2_pure_matter} and \eqref{eq:anomdim2_pure_matter} that satisfy the following relations
\begin{eqnarray}
	\left(\beta_{g_1} - \beta_{g_2} \right)|_{\mathcal{P}_{\mathbb{Z}_4}} = 0 \quad \textmd{and} \quad \left(\eta_1 -\eta_2 \right)|_{\mathcal{P}_{\mathbb{Z}_4}} = 0 \,,
\end{eqnarray}
and
\begin{eqnarray}
	\left(\beta_{g_1} - \beta_{g_2} \right)|_{\mathcal{P}_{\textmd{O}(2)}} = 0 \,, 
	\quad\!\! \left(\beta_{g_1} - \beta_{h_1} - 2\beta_{h_2}\right)|_{\mathcal{P}_{\textmd{O}(2)}} = 0 \,
	\quad\!\! \textmd{and} \quad\!\! \left(\eta_1 -\eta_2 \right)|_{\mathcal{P}_{\textmd{O}(2)}} = 0 \,,\,\,\,\,\,
\end{eqnarray}
where $\mathcal{P}_{\mathbb{Z}_4}$ and $\mathcal{P}_{\textmd{O}(2)}$ have been used to indicate that the expressions are projected into subspaces defined by \eqref{Z4_relations} and \eqref{O(2)_relations}, respectively. 

Although the results of the present section are restricted to the pure matter setting, the RG-consistence of symmetry-enhanced theory spaces remains valid in the presence of gravity. The case of scalars coupled to gravity is discussed in Sec.~\ref{sec:symmetries}.

\subsection{Multiple scalars in the $\textmd{O}(N_\phi)$ theory space}

In this section, we present results obtained within a model involving $N_\phi$ ($\geq2$) scalar fields with shift-symmetry and $\textmd{O}(N_\phi)$-symmetry. Our motivation is twofold: i) $\textmd{O}(N_\phi)$ symmetry is RG consistent even in the presence of gravity; ii) $\textmd{O}(N_\phi)$ symmetry limits the available interaction channels.

\begin{figure}[!t]
	\centering
	\hspace*{-0.5cm}\includegraphics[height=6cm]{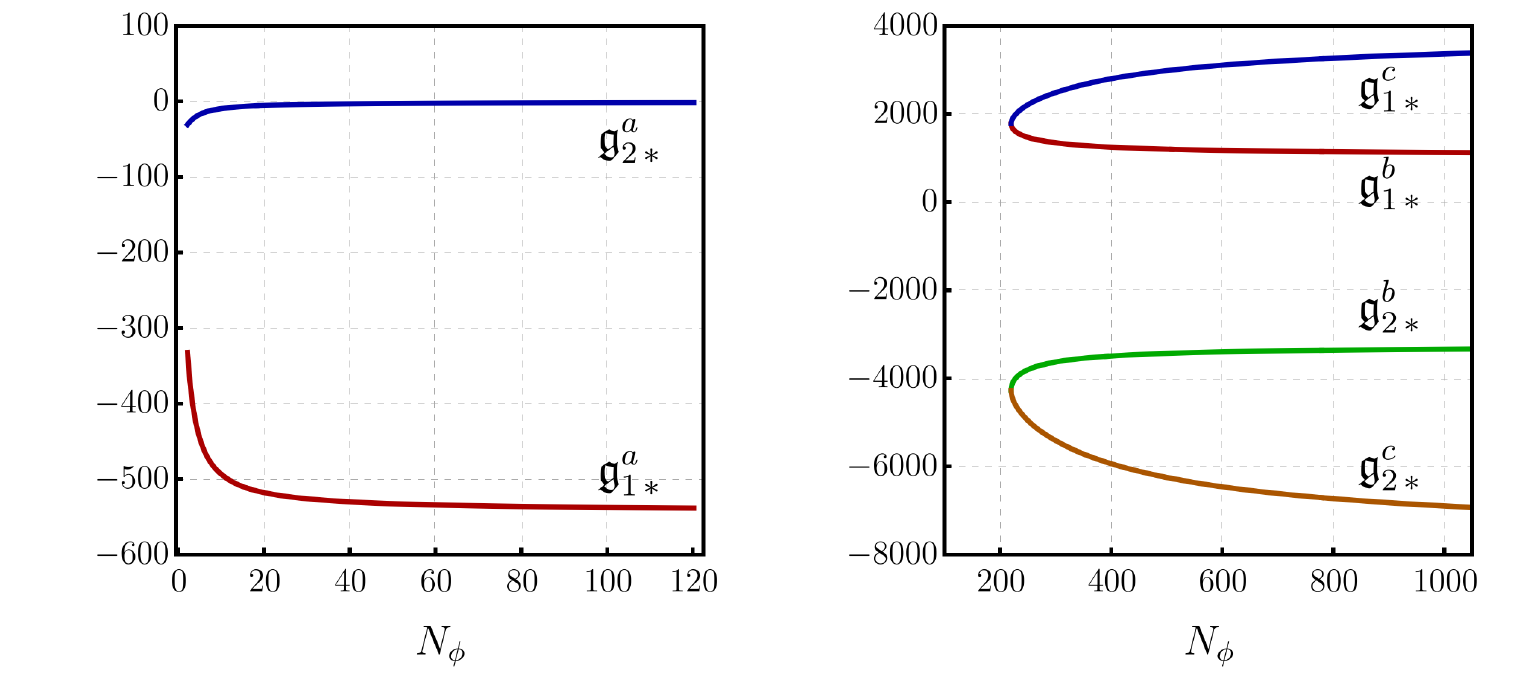}
	\caption{\label{fig:FPs_O(N)} Fixed point values in the theory space defined by shift- and $\textmd{O}(N_\phi)$-symmetries. The left-panel shows $\textmd{\textbf{FP}}_a$, while the right-panel correspond to the fixed point candidates $\textmd{\textbf{FP}}_b$ and $\textmd{\textbf{FP}}_c$.}
\end{figure}

\begin{figure}[!t]
	\centering
	\hspace*{-0.5cm}\includegraphics[height=6cm]{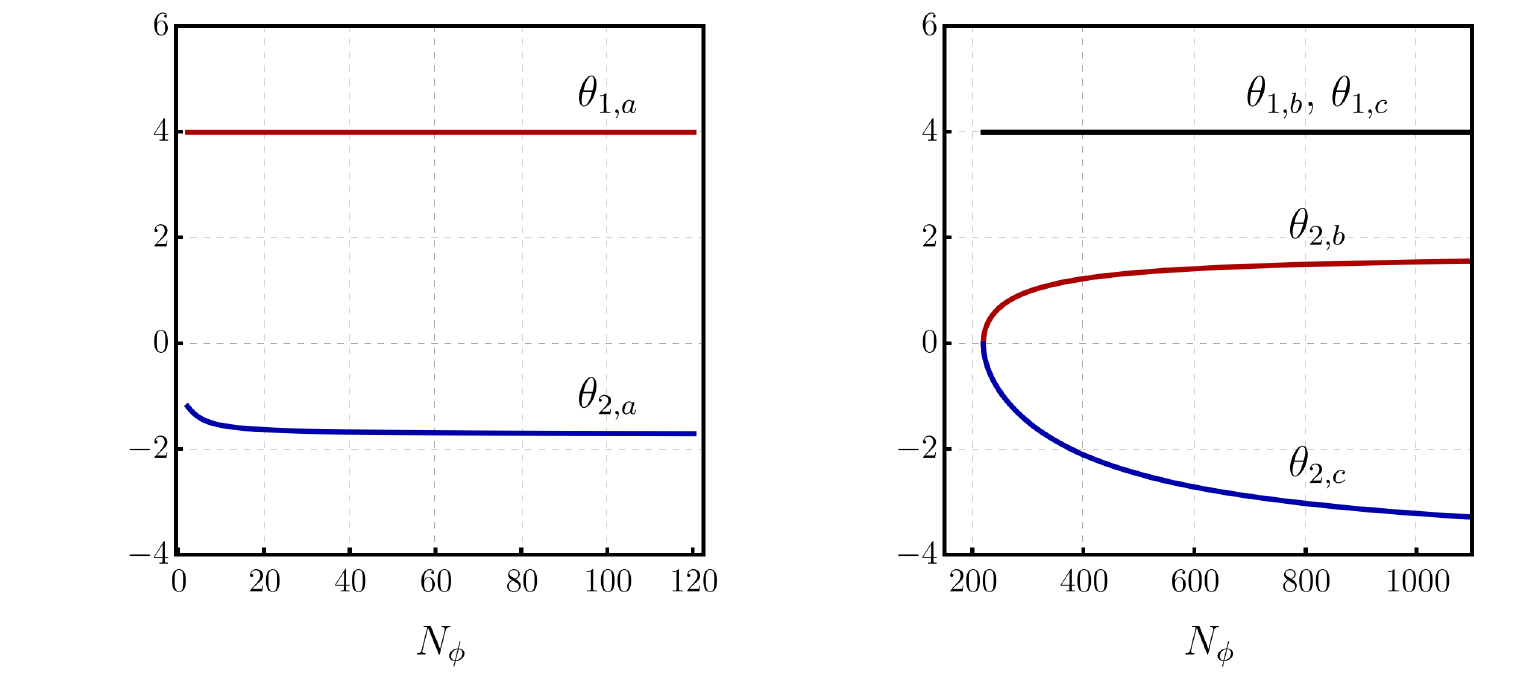}
	\caption{\label{fig:CritExp_O(N)} Critical exponents in the theory space defined by shift- and $\textmd{O}(N_\phi)$-symmetries. The left-panel shows the results corresponding to $\textmd{\textbf{FP}}_a$, while the right-panel correspond exhibits the critical exponents associated with $\textmd{\textbf{FP}}_b$ and $\textmd{\textbf{FP}}_c$.}
\end{figure}

\begin{figure}[!t]
	\centering
	\hspace*{-0.5cm}\includegraphics[height=6cm]{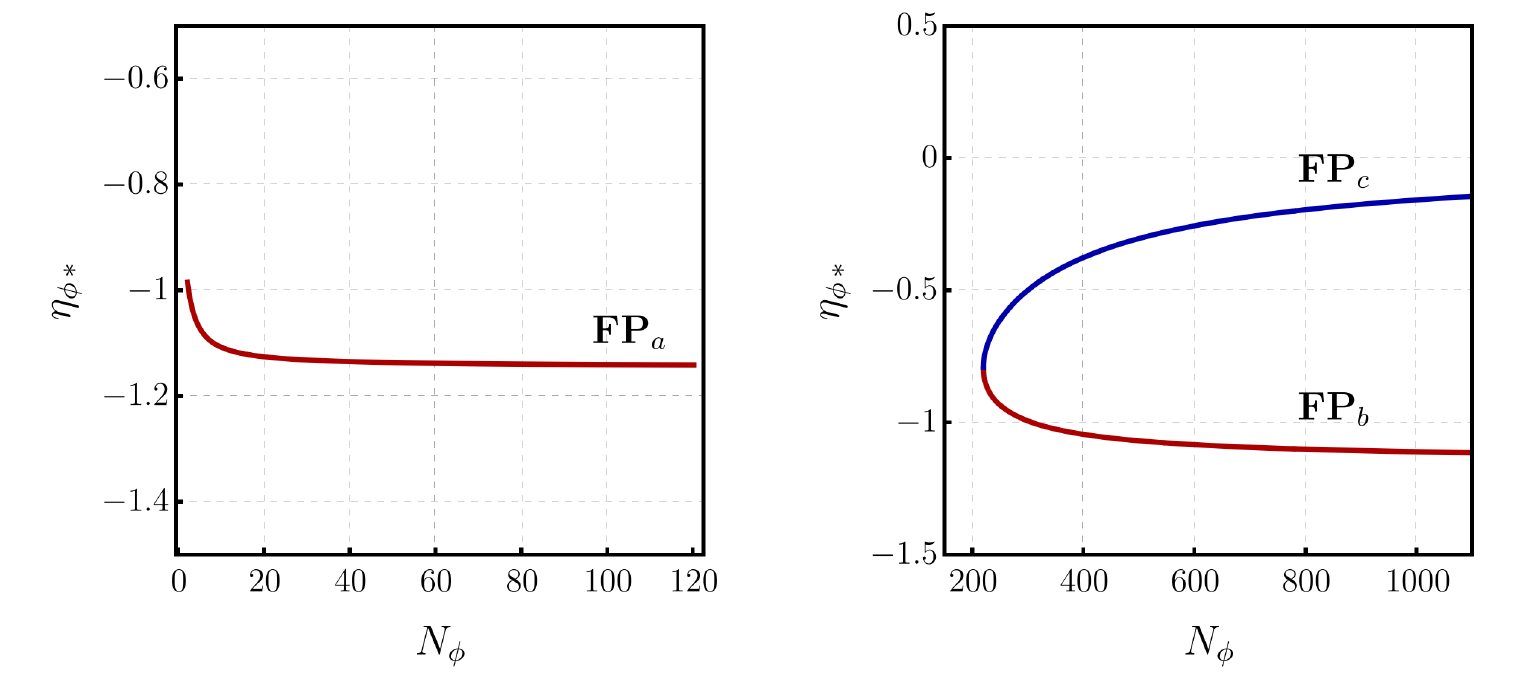}
	\caption{\label{fig:AnomDim_O(N)} Fixed point values for the scalar-field anomalous dimensions. The left (right) panel shows the results corresponding to $\textmd{\textbf{FP}}_a$ ($\textmd{\textbf{FP}}_b$ and $\textmd{\textbf{FP}}_c$).}
\end{figure}

Our truncation for the flowing action is given by \eqref{eq:Truncation_O(N)}. In the pure-matter setting, the resulting truncated beta functions are given by
\begin{eqnarray}\label{eq:betas_O(N)_PureMatter}
	\beta_{\mathfrak{g}_1} &=& (4+2\eta_\phi)\,\mathfrak{g}_1 + \frac{(6 N_\phi + 7)\,\mathfrak{g}_1^2 + (N_\phi + 15)\,\mathfrak{g}_2^2 + 2\,(3 N_\phi + 17) \mathfrak{g}_1	\mathfrak{g}_2 }{192\pi^2\, N_\phi} \, ,\\
	\beta_{\mathfrak{g}_2} &=& (4+2\eta_\phi)\,\mathfrak{g}_2  + \frac{\mathfrak{g}_1^2 +(N_\phi+21)\,\mathfrak{g}_2^2 + 10\, \mathfrak{g}_1	\mathfrak{g}_2 }{192\pi^2\, N_\phi} \,,
\end{eqnarray}
with anomalous dimension
\begin{eqnarray}\label{eq:etas_O(N)_PureMatter}
	\eta_ \phi = \frac{(2N_\phi+1)\mathfrak{g}_1 + (N_\phi + 5)\mathfrak{g}_2 }{96\pi^2\, N_\phi} \,.
\end{eqnarray}
Besides the Gaussian fixed point ($\mathfrak{g}_{1*} = \mathfrak{g}_{2*} = 0$), we have found three additional (non-Gaussian) fixed-point candidates \eqref{eq:betas_O(N)_PureMatter}, denoted as $\textmd{\textbf{FP}}_a$, $\textmd{\textbf{FP}}_b$ and $\textmd{\textbf{FP}}_c$. Although this set of fixed point candidates can be determined in an analytical way, the resulting expressions are too large and we shall not report them here. We show the fixed-point values as a function $N_\phi$ in Fig.~\ref{fig:FPs_O(N)} and critical exponents and anomalous dimensions in Figs.~\ref{fig:CritExp_O(N)} and \ref{fig:AnomDim_O(N)}.
%%%%%%%%%%%%

The fixed-point candidate $\textmd{\textbf{FP}}_a$ generalizes $\textmd{\textbf{FP}}_{3,1}$ (cf.~Tab.~\ref{tab:FPs_2scalars})  and  remains stable for any number of scalar fields (assuming $N_\phi \geq 2$) and has a single IR repulsive direction. The positive critical exponent is $\theta_{1,a} = 4$ for all values of $N_\phi$. The negative critical exponent varies slightly for a small number of scalar fields ($N_\phi \lesssim 20$) and stabilizes quickly at $ \theta_{2,a}\approx - 1.7$.

The fixed-point candidates $\textmd{\textbf{FP}}_b$ and $\textmd{\textbf{FP}}_c$ do not generalize a fixed-point candidate from the $N_{\phi}=2$ case because the quartic couplings are complex for $N_\phi \leq 218$. Both fixed-point candidates have a positive critical exponent with $\theta_{1,b} = \theta_{1,c} = 4$. The second critical exponent is positive for $\textmd{\textbf{FP}}_b$ and negative for  $\textmd{\textbf{FP}}_c$.

As can be inferred from Figs.~\ref{fig:FPs_O(N)}, \ref{fig:CritExp_O(N)} and \ref{fig:AnomDim_O(N)}, the fixed points approach an asymptotic behavior in the large-$N_\phi$. In this limit, the beta functions remain finite and are
\begin{eqnarray}\label{eq:betas_O(N)_PureMatter_LargeN}
	\beta_{\mathfrak{g}_1}|_{N_\phi \to \infty} &=& \left(4+2\eta_\phi|_{N_\phi \to \infty}\right)\mathfrak{g}_1 + \frac{6 \,\mathfrak{g}_1^2 + \mathfrak{g}_2^2 + 6 \,\mathfrak{g}_1	\mathfrak{g}_2 }{192\pi^2} \, ,\\
	\beta_{\mathfrak{g}_2}|_{N_\phi \to \infty} &=& \left(4+2\eta_\phi|_{N_\phi \to \infty}\right)\mathfrak{g}_2  + \frac{\mathfrak{g}_2^2}{192\pi^2} \,,
\end{eqnarray}
with
\begin{eqnarray}\label{eq:etas_O(N)_PureMatter_LargeN}
	\eta_ \phi|_{N_\phi \to \infty} = \frac{2\,\mathfrak{g}_1 + \mathfrak{g}_2 }{96\pi^2} \,.
\end{eqnarray}
In this regime, the fixed point candidates converge to the results reported in Tab.~\ref{tab:FPs_O(N)-Large_N}.
\begin{table}[t]
	\begin{center}
		\begin{tabular}{|c|c|c|c|c|c|}
			\hline\hline 
			& $\mathfrak{g}_{1*}$ & $\mathfrak{g}_{2*}$  & $\eta_{\phi\,*}$  & $\theta_1$ & $\theta_2$ \\ \hline\hline
			$\textmd{\textbf{FP}}_{a}|_{N_\phi \to \infty} $ & $-384\pi^2/7$ & $0$   & $-8/7$ & $4$ & $-12/7$  \\
			$\textmd{\textbf{FP}}_{b}|_{N_\phi \to \infty} $ & $768\pi^2/7$ & $-2304 \pi ^2/7$   & $-8/7$ & $4$ & $12/7$  \\
			$\textmd{\textbf{FP}}_{c}|_{N_\phi \to \infty} $ & $384\pi^2$ & $-768 \pi ^2$ &  $0$ & $4$ & $-4$  \\\hline\hline
		\end{tabular}
		\caption{Fixed-point candidates in the large-$N_\phi$ limit of a scalar field model defined in terms of shift- and $\textmd{O}(N_\phi)$-symmetry. In addition to fixed-point values, we report the corresponding anomalous dimensions and critical exponents.}
		\label{tab:FPs_O(N)-Large_N}
	\end{center}
\end{table}

\subsection{Fixed point candidates away from four-dimensions}
A prominent mechanism for asymptotic safety is to start from the free fixed point in the critical dimension of an interaction, where this interaction has a dimensionless coupling. When one
changes the spacetime dimensionality away from the critical dimension $d_c$, the canonical scaling and quantum scaling balances out to generate an interacting fixed point. This mechanism is at work in, e.g., Yang-Mills theory in $d=4+\epsilon$ \cite{Gies:2003ic,Morris:2004mg,Florio:2021uoz}, the non-linear sigma model in $d=2+\epsilon$ \cite{Codello:2008qq} and in quantum gravity in $d=2+\epsilon$ \cite{Christensen:1978sc,Kawai:1992np,Martini:2021slj}. For sufficiently small $\epsilon$, the one-loop term in the beta function is sufficient to characterize the interacting fixed point. Schematically, for a coupling $\bar{\gamma}$ that is dimensionless in $d_{\rm crit}$, has mass-dimension $-\epsilon$ in $d=d_{\rm crit}+ \epsilon$, and a one-loop term $\mathcal{O}(\gamma^2)$, the beta function for the dimensionless coupling $\gamma = \bar{\gamma}\, k^{\epsilon}$ is given by
\begin{eqnarray}
	\beta_{\gamma} = \epsilon\, \gamma + \beta_1\, \gamma^2 + \mathcal{O}(\gamma^3).
\end{eqnarray}
For sufficiently small $\epsilon$, the interacting fixed point at
\begin{eqnarray}
	\gamma_{\ast} = -\, \epsilon/\beta_1,
\end{eqnarray}
is well under control in perturbation theory.
For larger $\epsilon$, the $\epsilon$ expansion must be supplemented by resummation techniques. Alternatively, the functional Renormalization Group can be used to continuously interpolate between dimensionalities, while an appropriate choice of truncation allows to control the theory also beyond the perturbative regime. Conversely, the reliability of a fixed-point candidate in $d \neq d_{\rm crit}$ can be checked by following it continuously while changing $d$ towards $d_{\rm crit}$. In this way, a deformation of a fixed point may be found that is under control in perturbation theory.

Accordingly, we use the FRG to explore shift-symmetric scalar theories between $d_{\rm crit}$ and $d=4$. There is an additional subtlety not present in the above examples, since $d_{\rm crit} =0$ for the shift-symmetric interactions. The zero-dimensional limit of a QFT is a purely mathematical limit, in which no degrees of freedom exist. Similar limits of QFTs may nevertheless encode physics, such as the O$(N)$ model in three dimensions, where the limit $N \rightarrow 0$ does not feature any degrees of freedom, but nevertheless encodes the universality class of self-avoiding random walks \cite{deGennes:1972zz}.

Within the setting discussed here, we can continuously interpolate between the free fixed point in $d = d_{\rm crit} = 0$ and an interacting fixed-point candidate in $d=4$. In the simple shift-symmetric model containing a single scalar-field (i.e., the $d$-dimensional version of \eqref{eq:truncationsinglescalar} without gravity), the flow of the quartic coupling $g$ is given by
\begin{eqnarray}
	\beta_g = d \,g + \frac{d^2 + 6\,d + 14}{(4\pi)^{d/2}\,\Gamma(d/2+3)} \,g^2 \,,
\end{eqnarray}
leading to the interacting fixed point
\begin{eqnarray}\label{eq:FP_d-dim_SingleScalar}
	g_* = -\frac{(4\pi)^{d/2}\,d\,\Gamma(d/2+3)}{d^2 + 6\,d + 14} .
\end{eqnarray} 
The corresponding critical exponent is $\theta(g^*) = d$. Eq.~\eqref{eq:FP_d-dim_SingleScalar}, shows that the interacting fixed point $g_*= -256\pi^2/9$ in four-dimensions can be continuously connected to the near-Gaussian fixed point $g_* = - \epsilon/7 + \mathcal{O}(\epsilon^2)$ in $d=\epsilon \ll 1$. The latter has a near-canonical critical exponent $\theta(g^*)|_{d=\epsilon} = \epsilon$. This follow directly from the fact that at the free fixed point, $g$  is irrelevant for $d>4$, and thus it must be relevant at the interacting one for the single-field model, as otherwise the beta function would contain a discontinuity or singularity. Enlarging the theory space to the case with many scalar-fields, not all shift-symmetric couplings must become relevant in order to enable an interacting fixed point; in fact, a single relevant direction suffices.

\begin{figure}[!t]
	\centering
	\hspace*{-0.5cm}\includegraphics[height=6cm]{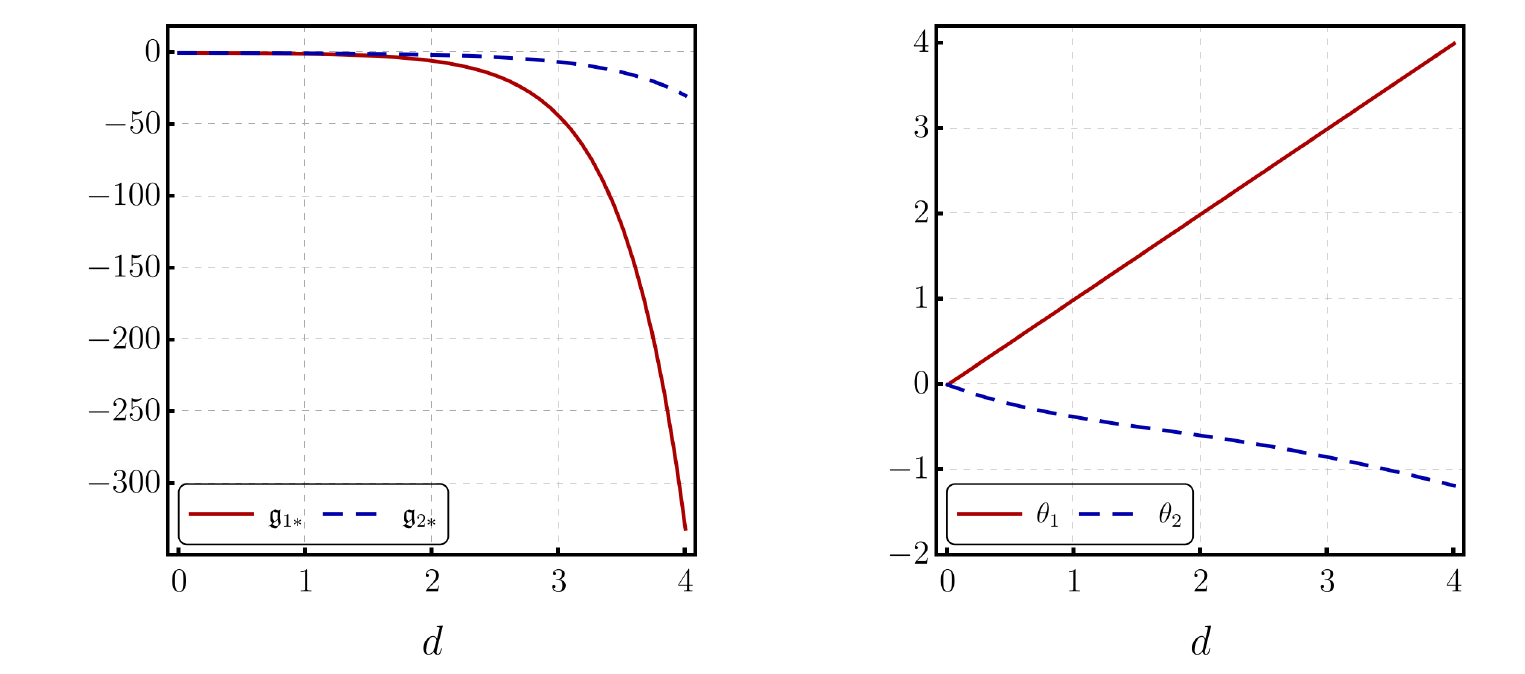}
	\caption{\label{fig:connecting_to_d=0} We show fixed-point candidates (left panel) and the corresponding critical exponents (right panel) as a functions of dimensionality. This result was obtained with truncation featuring both O$(2)$- and shift-symmetry in $d$-dimensions.}
\end{figure}

In Fig.~\ref{fig:connecting_to_d=0} we show the fixed points and corresponding critical exponents in a $d$-dimensional version of the O$(N_\phi)$-model discussed in the previous section. We report only the results for $N_\phi = 2$, but the qualitative behavior remains the same for other choices of $N_\phi$. A continuous interpolation from a perturbative fixed point in $d=\epsilon$ to the interacting fixed-point candidates in Tab.~\ref{tab:FPs_2scalars}  in $d=4$ is always possible.
Going beyond fixed points with O$(2)$ symmetry, we verified that the other interacting fixed points reported in Table \ref{tab:FPs_2scalars} can also be continuously connected to perturbative fixed points when we approach $d=0$.

\section{Gauge dependence}\label{app:GaugeDep}

The results presented in the main text were obtained with gauge choice $\alpha \to 0$ and $\beta \to 0$. For the gauge parameter $\alpha$, \cite{Knorr:2017fus} shows that $\alpha \to 0$ has the status of fixed point solution and, therefore, it can be considered a preferred gauge choice for $\alpha$. For the gauge parameter $\beta$, \cite{Knorr:2017fus} shows that $\beta$ does not flow in the limit $\alpha \to 0$ and, therefore, we do not have a preferred gauge choice for $\beta$.  In this appendix we investigate how different choices of $\beta$ affect our results.

For $\beta \neq 0$, the graviton propagator involves the TT-mode and a mixture of the scalar modes $\sigma$ and $h$, cf.~Eq.~\eqref{eq:YorkDecomp}. With this mixture of the scalar modes, it is useful to express the graviton propagator using the basis of projectors introduced in \cite{Knorr:2021niv}.

For $N_\phi = 1$ and $\Lambda = 0$, the $\beta$-dependent beta function for the coupling $g$ is given by
\begin{eqnarray}
	\beta_g &=& 4 g + \frac{9 g^2}{64\pi^2} - \left( \frac{5}{6 \pi } + \frac{22}{3 \pi  (3-\beta )} - \frac{6}{\pi  (3-\beta )^2} \right)\!gG \nonumber\\
	&+&\left( 72-\frac{64}{3-\beta }+\frac{576}{(3-\beta )^2}-\frac{1536}{(3-\beta )^3}+\frac{1152}{(3-\beta )^4} \right) G^2 \,.
\end{eqnarray}
From the above equation we can see that the flow of $g$ is well defined for $\beta \neq 3$. The choice $\beta = 3$ is a pole in the integrated graviton propagator, which results from an incomplete implementation of the gauge-fixing term. We concentrate our analysis to the region $\beta < 3$, where we can continuously connect different gauge choices to $\beta = 0$.

\begin{figure}[!t]
	\centering
	\hspace*{-1.cm}\includegraphics[height=7.0cm]{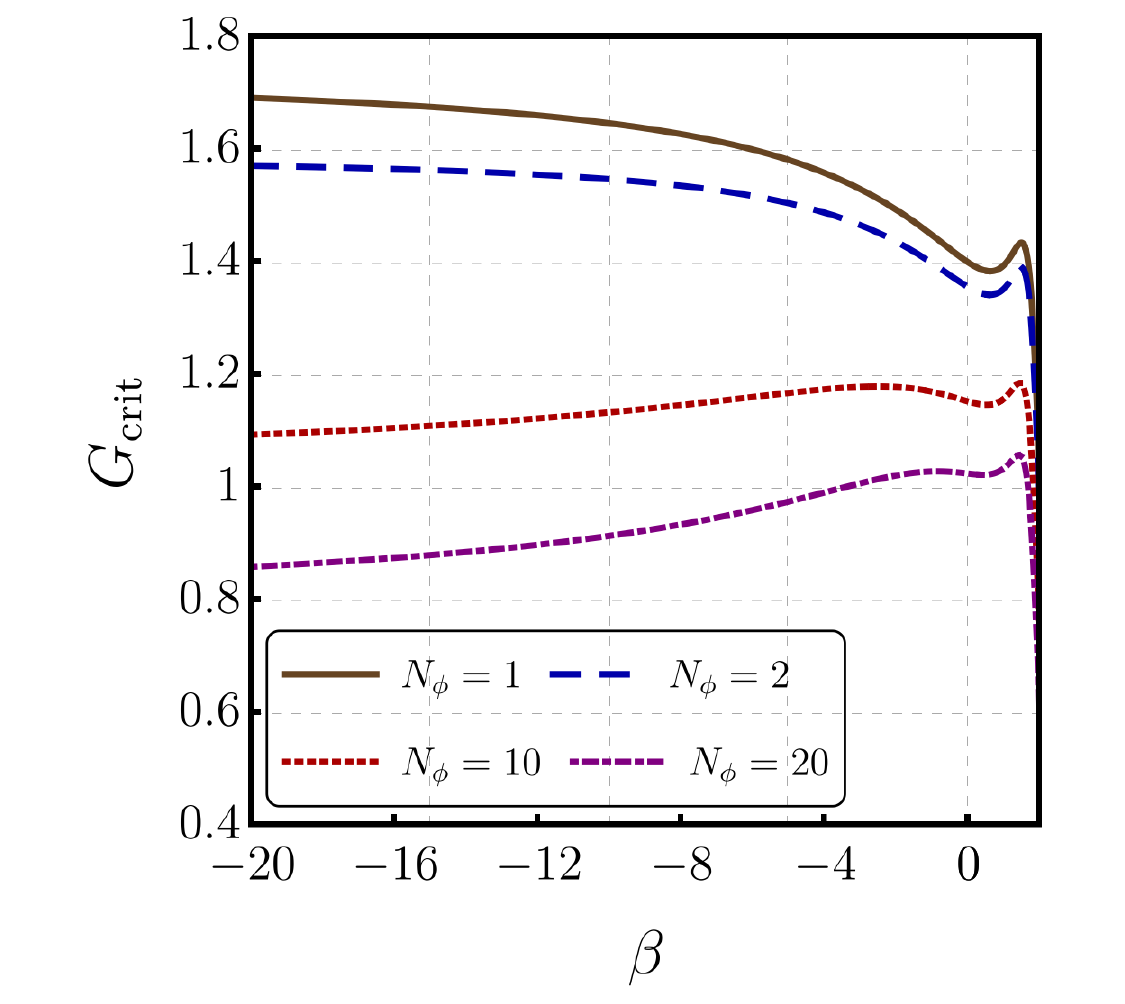}
	\caption{\label{fig:GaugeDep_Lambda=0} In this plot we show $G_{\textmd{crit}}$ as a function of the gauge parameter $\beta$. The different curves correspond to different values of $N_\phi$.}
\end{figure}

For $N_\phi = 1$ and $\Lambda = 0$, the $\beta$-dependent critical value $G$ separating the weak-gravity from the strong-gravity regime is given by
\begin{eqnarray}
	G_\textmd{crit}(\beta) = \frac{24 \pi  (\beta -3)^2}{5 \beta ^2 -74 \beta + 141 + 9 \sqrt{2} \sqrt{9 \beta ^4 - 100 \beta ^3 + 486 \beta ^2 - 996 \beta + 729}}\,.
\end{eqnarray}
Despite the apparently complicated $\beta$-dependence of $G_\textmd{crit}$, Fig.~\ref{fig:GaugeDep_Lambda=0} shows a mild gauge dependence for values of $\beta \lesssim 1.5$. The region between $1.5 \lesssim \beta < 3$ show a strong variation on $G_\textmd{crit}$. However, this region should be excluded from our analysis due to the proximity to the pole at $\beta=3$. Fig.~\ref{fig:GaugeDep_Lambda=0} shows a similar behavior for other choices of $N_\phi$.

\begin{figure}[!t]
	\centering
	\hspace*{-.1cm}\includegraphics[height=4.5cm]{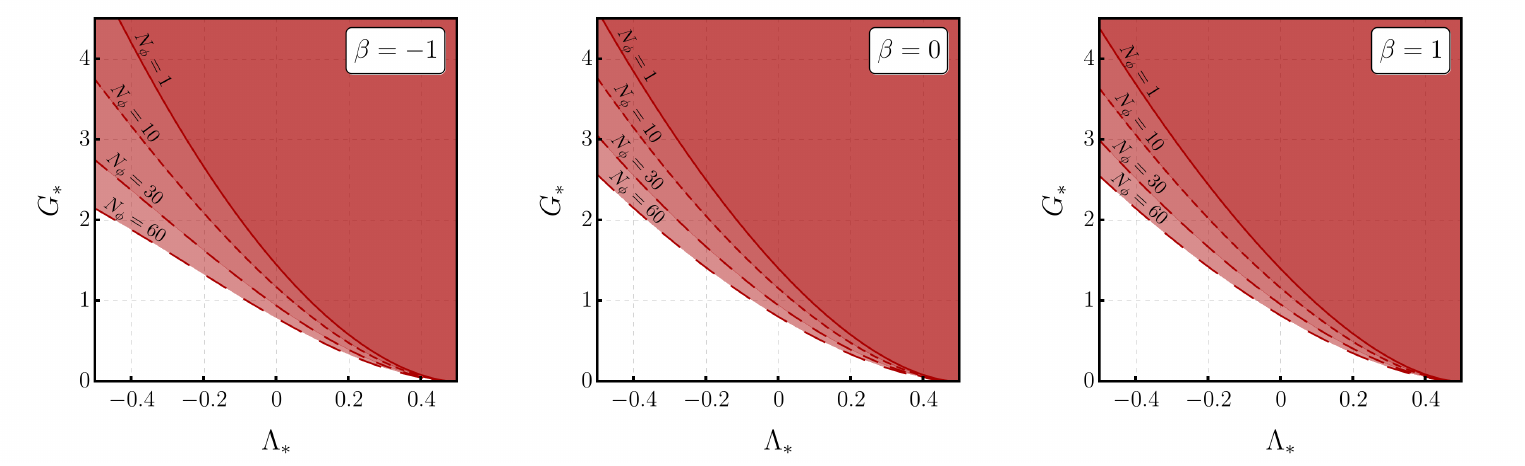}
	\caption{\label{fig:GaugeDep_Lambda-Gplane}
		In this plot we show the weak-gravity bound for multiple values of $N_{\phi}$. Each panel corresponds to a particular choice of the gauge parameter $\beta$. }
\end{figure}

For $\Lambda \neq 0$, each choice of the gauge parameter $\beta$ defines a regions excluded by the weak-gravity bound in the $G - \Lambda$ plane. In Fig.~\ref{fig:GaugeDep_Lambda-Gplane} we show the weak-gravity bound for three choices of $\beta$ and for multiple values of $N_\phi$. Fig.~\ref{fig:GaugeDep_Lambda-Gplane}  indicates that different choices of $\beta$ does not affect the qualitative features of our results.  

\section{Strengthening the evidence for the prediction of Higgs and top quark masses}\label{app:irrelevance}
In this appendix we consider the contribution of $\bar{g}(\del_\mu \phi)^4$, see Sec.~\ref{sec:WGB1scalar}, to beta functions for a quartic scalar and a Yukawa coupling. The values of these canonically marginal couplings may be predicted from the asymptotic safety paradigm. In \cite{Shaposhnikov:2009pv}, a scenario was suggested in which the Higgs quartic coupling is irrelevant at a gravity-induced free fixed point, resulting in a prediction of the Higgs mass. Indications for this scenario were then found, e.g., in \cite{Oda:2015sma,Wetterich:2016uxm,Eichhorn:2017als,Pawlowski:2018ixd,Wetterich:2019rsn,Eichhorn:2020sbo}. Yukawa couplings may correspond to predictions of the asymptotic-safety paradigm at a gravity-induced interacting fixed point \cite{Eichhorn:2017eht}, potentially resulting in a calculable top-quark and bottom-quark mass \cite{Eichhorn:2017ylw, Eichhorn:2018whv, Alkofer:2020vtb}.

\subsection{Higgs quartic coupling}

\begin{figure}[!t]
	\centering
	\hspace*{-.75cm}\includegraphics[height=6.5cm]{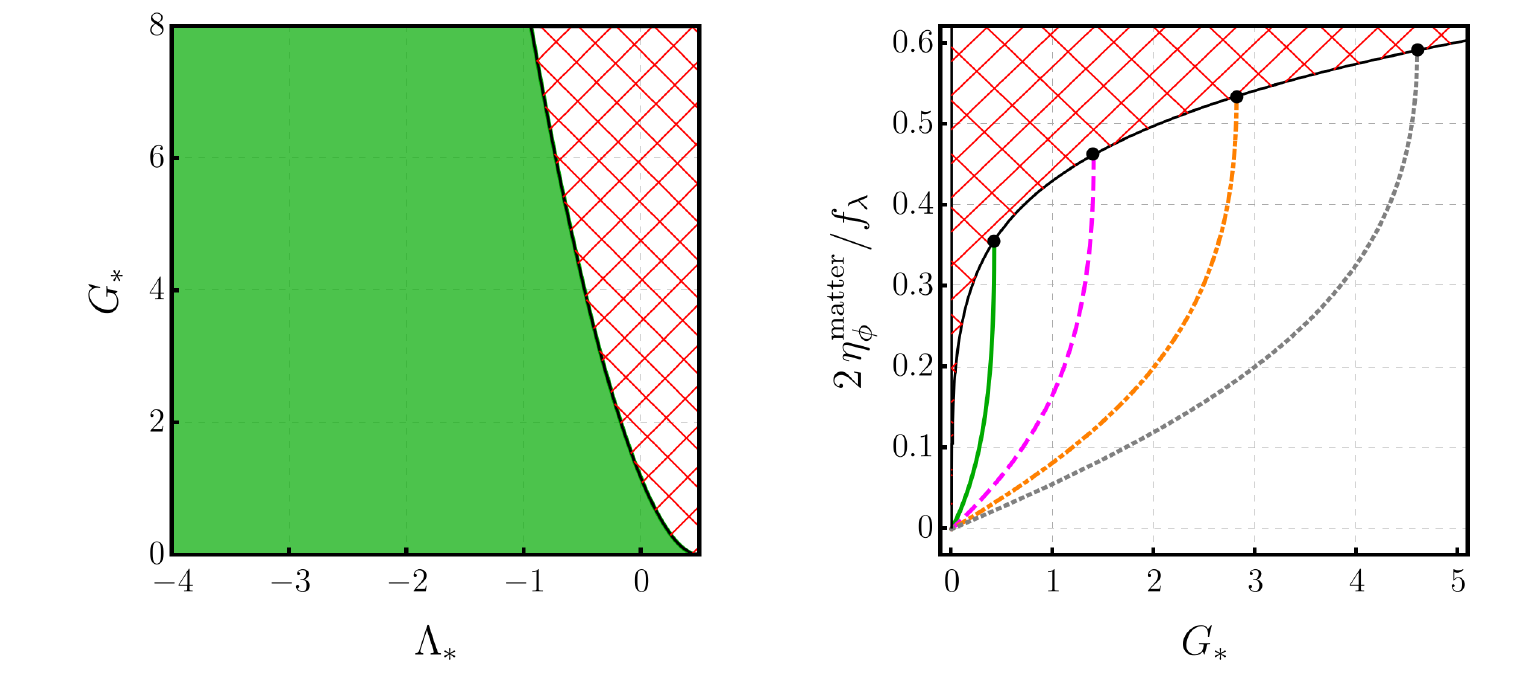}
	\caption{\label{fig:fHiggs} Left panel: The green part corresponds to the region where $F_{\lambda}(G_{\ast}, \Lambda_{\ast}, g_{\ast})$ is negative, and coincides with the constraint from the weak-gravity bound. 
	In the green region, $f_{\lambda}(G_{\ast}, \Lambda_{\ast})<0$ and $\eta_{\phi}^\textmd{matter}(g_*)<0$. 
	Right panel: The ratio between $\eta_{\phi}^\textmd{matter}(g_*)$ and $f_{\lambda}$ is small, showing that the matter contribution is subleading. The ratio is plotted as a function of $G_*$, for different values of $\Lambda_{*}$, namely, $\Lambda^{(1)}_{*}=0.25$ (green continuous line), $\Lambda^{(2)}_{*}=0$ (magenta dashed line),  $\Lambda^{(3)}_{*}=-0.25$ (orange dot-dashed line), and  $\Lambda^{(4)}_{*}=-0.5$ (gray dotted line).  }
\end{figure}

We consider the quartic coupling in a $\lambda\phi^4$ model. The Higgs sector of the Standard Model contains a complex SU(2) scalar with its corresponding quartic interaction. The gravitational contribution to the two quartic couplings is universal; thus we focus on the real scalar here. To leading order, its beta function is given by
\begin{eqnarray}
	\beta_\lambda= - F_{\lambda}(G, \Lambda, g) \lambda + \mathcal{O}	(\lambda^2).
\end{eqnarray}
Herein, $F_{\lambda}(G, \Lambda, g)$ consists of a contribution that has been calculated previously and a contribution that we are calculating for the first time here:
\begin{eqnarray}
	F_{\lambda}(G, \Lambda, g) = f_{\lambda}(G, \Lambda) - 2\,\eta_{\phi}^\textmd{matter}(g).
\end{eqnarray}
The direct gravitational contribution $f_{\lambda}(G_{\ast}, \Lambda_{\ast})$ depends on the gravitational fixed-point values and has been calculated in the Einstein-Hilbert truncation in \cite{Narain:2009fy,Eichhorn:2017als,Pawlowski:2018ixd,Eichhorn:2020sbo} and beyond in \cite{Hamada:2017rvn,deBrito:2019umw}. For our choice of gauge and regulator, it reads
\begin{eqnarray}
	f_{\lambda}(G,\Lambda) = -\frac{5 G}{2 \pi  (1-2 \Lambda )^2} - \frac{G}{9 \pi  \left(1-4 \Lambda /3\right)} - \frac{4 G}{9 \pi  \left(1-4 \Lambda /3\right)^2} \,.
\end{eqnarray}
The second contribution, $\eta_{\phi}^\textmd{matter}(g)$ is an \emph{indirect} gravitational contribution: It is given in terms of a pure-scalar tadpole diagram, which is $\sim g$. Under the impact of gravity, $g_{\ast}\neq 0$, such that this additional contribution is present. Here, we will explore whether this additional contribution can change the sign of $F_{\lambda}(G_{\ast}, \Lambda_{\ast}, g_{\ast})$, compared to $f_{\lambda}(G_{\ast}, \Lambda_{\ast})$. The physical significance of the question lies in the prediction of the Higgs mass: if $\lambda$ is irrelevant at the fixed point $\lambda_{\ast}=0$, the ratio of Higgs mass to the electroweak vacuum expectation value is calculable from asymptotic safety, as proposed in \cite{Shaposhnikov:2009pv}, see also \cite{Wetterich:2019qzx}. The critical exponent around the Gaussian fixed point is given by $F_{\lambda}(G_{\ast}, \Lambda_{\ast}, g_{\ast})$. If it is positive,  then the coupling is relevant and the fixed point is IR repulsive; if it is negative, then the coupling is irrelevant and the fixed point is IR attractive. 
In fact, $f_{\lambda}(G_{\ast}, \Lambda_{\ast})<0$ for any $G_{\ast}>0$ and $\Lambda_{\ast} < 1/2$, cf.~Fig.~\ref{fig:fHiggs}.

However, the matter sector of our model - quartic in derivatives - contributes to $\eta_{\phi}^\textmd{matter}(g)$. This contribution might change the irrelevant character of the coupling $\lambda$. The same might be true for higher derivative terms beyond our truncation. Nonetheless, we would expect that the contribution of these derivative terms that together make up $\eta_{\phi}^\textmd{matter}(g)$ are not large enough to render the coupling relevant. The basis for this expectation is that parametrically, the matter couplings that enter $\eta_{\phi}^\textmd{matter}(g)$ are set by the gravitational fixed-point value $G_{\ast}$ divided by a loop factor $16\pi^2$. Accordingly, we expect them to be subleading compared to the direct gravitational contribution in $f_{\lambda}(G_{\ast}, \Lambda_{\ast})$, see also \cite{Eichhorn:2017eht}. We find that
\begin{eqnarray}
	\eta_{\phi}^\textmd{matter}(g_*) = \dfrac{g_*}{32\pi^2}\,, \label{eq:mattercontr}
\end{eqnarray}
where $g_{\ast}$ assumes negatives values, cf.~Fig.~\ref{fig:FPcollision}, so that it might change the irrelevant character of the quartic coupling. However, the matter contribution is not large enough to drive the  quartic coupling towards relevance, cf.~Fig.~\ref{fig:fHiggs}. As outlined above, we expect the same subleading character to hold for higher-order matter interactions; $g$ serves as an explicit example to demonstrate that induced matter couplings are subleading compared to the direct gravitational contribution to the critical exponent of the Higgs quartic coupling. We thus conclude that the proposal for the Higgs-mass prediction advanced in \cite{Shaposhnikov:2009pv} passes another consistency test and remains viable under the present extension of the truncation.

\subsection{Yukawa coupling}

\begin{figure}[!t]
	\centering
	\hspace*{-1cm}\includegraphics[height=6.5cm]{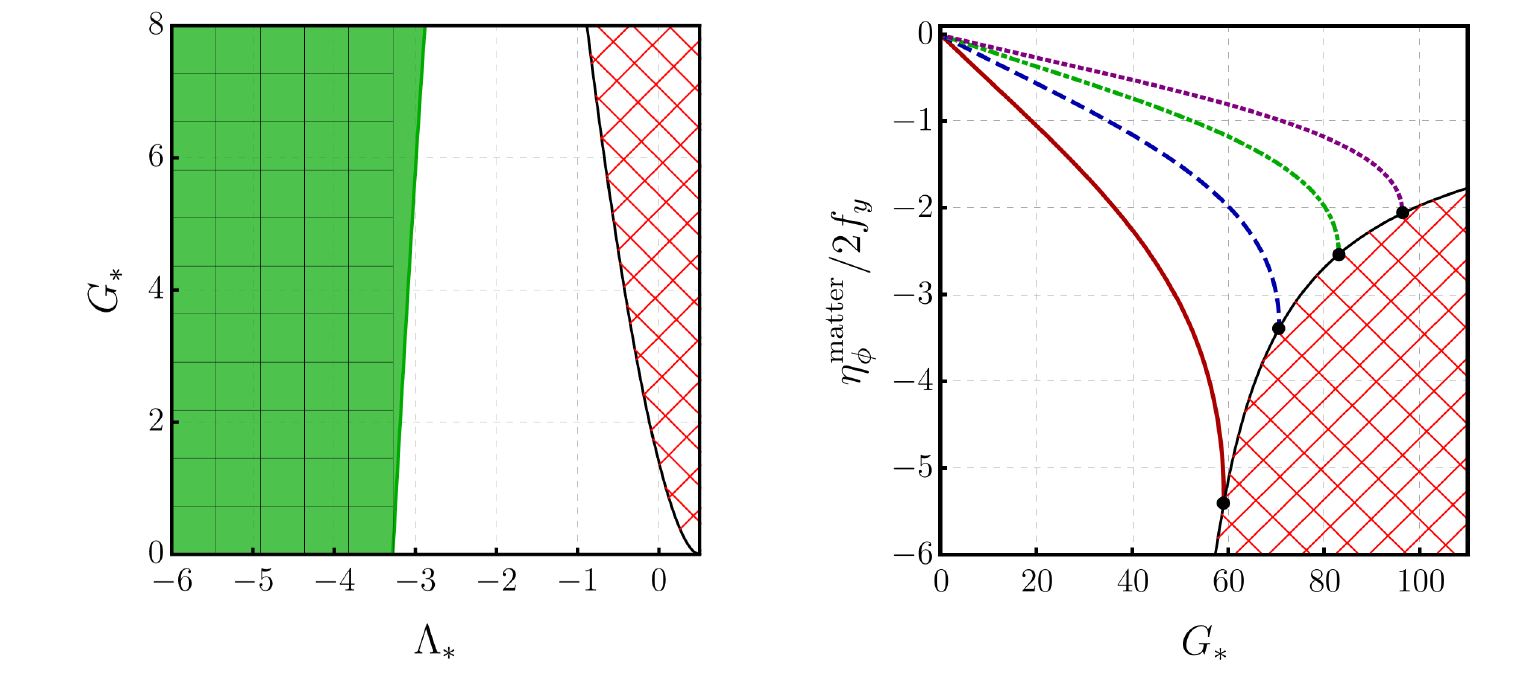}
	\caption{\label{fig:regionplotfYukawa} 
	Left panel: The green part corresponds to the region where $F_y(G_{\ast},\Lambda_{\ast},g_{\ast})>0$. Within the region indicated by gray squares, $f_y(G_{\ast},\Lambda_{\ast})>0$. Including the derivative interaction increases the allowed region for $G_{\ast}$ and $\Lambda_{\ast}$. The red-white region is the one excluded by the weak-gravity bound.  
	Right panel: The ratio between $\eta_{\phi}^\textmd{matter}(g_*)$ and $f_{y}$ is plotted as a function of $G_*$, for different values of $\Lambda_{*}$, namely, $\Lambda_{*}=-4$ (dark red continuous line), $\Lambda_{*}=-4.5$ (blue dashed line),  $\Lambda_{*}=-5$ (green dot-dashed line), and  $\Lambda_{*}=-5.5$ (purple dotted line).
	}
\end{figure}

In contrast to the case of the Higgs quartic coupling, a prediction of a nonzero top quark mass relies on the existence of a non-Gaussian fixed point. If the Yukawa coupling was irrelevant at a Gaussian fixed point, it would vanish at all scales -- in contrast to the Higgs quartic coupling, which is regenerated by quark and gauge boson fluctuations below the Planck scale. A vanishing Yukawa coupling is not compatible with the experimental data from the LHC \cite{CMS:2019art,Agaras:2020zvy}. For the top-Yukawa coupling $y$, the beta function with gravity is given by \cite{Eichhorn:2017ylw}
\begin{eqnarray}
	\beta_y=-F_y(G,\Lambda,g)\,y+\dfrac{9}{32\pi^2} y^3,
\end{eqnarray}
such that
\begin{eqnarray}
	F_y(G_{\ast}, \Lambda_{\ast}, g_{\ast}) = f_y(G_{\ast}, \Lambda_{\ast}) -\dfrac{1}{2}\eta^\textmd{matter}_\phi(g_*).
\end{eqnarray}
As in the case of the gauge coupling, the gravitational contribution $F_y$ consists of an indirect gravitational contribution $\eta_{\phi}^\textmd{matter}(g_{\ast})$, given by \eqref{eq:mattercontr}, and of a direct gravitational contribution $f_y$, computed in \cite{Oda:2015sma,Eichhorn:2016esv,Eichhorn:2017ylw}. For our choice of gauge and regulator,
\begin{eqnarray}
	f_y(G,\Lambda) = -\frac{15 G}{16 \pi  (1-2 \Lambda )^2}+\frac{7 G}{72 \pi  \left( 1-4\Lambda/3 \right)} - \frac{7 G}{144 \pi  \left(1-4\Lambda/3 \right)^2} \, . \label{eq:fy}
\end{eqnarray}
The existence of a nontrivial IR-attractive fixed point at $y_*\neq 0$ demands that $f_y(G_*,\Lambda_*)>0$. This is equivalent to demanding that the Gaussian fixed point is IR repulsive. At the free fixed point, $F_y$ corresponds to the critical exponent and thereby constitutes a universal quantity. 

From \eqref{eq:fy}, $f_y(G_*,\Lambda_*)$ is zero at a critical value $\Lambda_{\ast,\textmd{crit}}\approx -3.28$, and it is positive for $\Lambda_{\ast}<\Lambda_{\ast,\textmd{crit}}$, cf.~Fig.~\ref{fig:regionplotfYukawa}.
Since the matter contribution for $\eta^\textmd{matter}_\phi$ is negative, the presence of derivative interactions slightly increases the viable parameter space for $(G_{\ast},\Lambda_{\ast})$. The matter contribution is small but not always subleading, because $f_y=0$ when $\Lambda_{\ast}=\Lambda_{\ast,\textmd{crit}}$.

\bibliography{references}
\end{document}